\documentclass[groupedaddress,
 reprint,
 amsmath,amssymb,
 nofootinbib,
 superscriptaddress,
 aps,
 prd,
]{revtex4-1}

\usepackage{dcolumn}
\usepackage{bm}
\usepackage{hyperref}
\usepackage[utf8]{inputenc}
\usepackage[capitalise]{cleveref}
\usepackage{aas_macros}
\usepackage[normalem]{ulem}
\usepackage{xcolor}
\usepackage{csquotes}
\usepackage{graphicx}

\usepackage[makeroom]{cancel}
\usepackage{printlen}

\usepackage{mathtools}
\usepackage{physics}

\usepackage{amsmath}

\makeatletter
\newcommand*\rel@kern[1]{\kern#1\dimexpr\macc@kerna}
\newcommand*\widebar[1]{%
  \begingroup
  \def\mathaccent##1##2{%
    \rel@kern{0.8}%
    \overline{\rel@kern{-0.8}\macc@nucleus\rel@kern{0.2}}%
    \rel@kern{-0.2}%
  }%
  \macc@depth\@ne
  \let\math@bgroup\@empty \let\math@egroup\macc@set@skewchar
  \mathsurround\z@ \frozen@everymath{\mathgroup\macc@group\relax}%
  \macc@set@skewchar\relax
  \let\mathaccentV\macc@nested@a
  \macc@nested@a\relax111{#1}%
  \endgroup
}
\makeatother

\begin{document}

\title{Post-inflationary structure formation boosted by parametric self-resonance}

\author{Benedikt Eggemeier}
\email{benedikt.eggemeier@phys.uni-goettingen.de}
\affiliation{
 Institut f\"ur Astrophysik, Georg-August-Universit\"at G\"ottingen, D-37077 G\"ottingen, Germany
}

\author{Peter Hayman}
\email{peter.hayman@auckland.ac.nz}
\affiliation{Department of Physics, University of Auckland, Private Bag 92019, Auckland, New Zealand}

\author{Jens C. Niemeyer}
\email{jens.niemeyer@phys.uni-goettingen.de}
\affiliation{
 Institut f\"ur Astrophysik, Georg-August-Universit\"at G\"ottingen, D-37077 G\"ottingen, Germany
}

\author{Richard Easther}
\email{r.easther@auckland.ac.nz}
\affiliation{Department of Physics, University of Auckland, Private Bag 92019, Auckland, New Zealand}

\date{\today}

\begin{abstract}
The post-inflationary Universe can pass through a long epoch of effective matter-dominated expansion. This era may allow for both the parametric amplification of initial fluctuations and the gravitational collapse of inflaton perturbations.  
We perform first-of-their-kind high-resolution simulations that span the resonant phase and the subsequent gravitational collapse of the inflaton field by seguing from a full Klein-Gordon treatment of  resonance to a computationally efficient Schrödinger-Poisson description that accurately captures the gravitational dynamics when most quanta are nonrelativistic. We consider a representative example in which resonance generates $\mathcal{O}(10^{-1})$ overdensities and gravitational collapse  follows promptly as resonance ends.  We observe the formation of solitonic cores inside inflaton halos and complex gravitational dynamics on scales of $10^{-27}\,\mathrm{m}$, greatly extending the possible scope of nonlinear post-inflationary gravitational dynamics. 
\end{abstract}

\maketitle

\section{Introduction}

In early universe inflationary scenarios~\cite{Starobinsky1980,Guth1981,Linde1982,Linde1983}, the energy density in the inflaton field must eventually be transformed into Standard Model particles and dark matter.  This process is known as reheating and is not well constrained by observations, leaving room for a plethora of different mechanisms to thermalize the Universe. In many models inflation ends near Grand Unification energies but reheating need not be completed until temperatures are at the MeV scale~\cite{Kawasaki1999,Hannestad2004, Salas2015}. Consequently, the Universe can grow substantially before the onset of the radiation domination.

In single-field models the inflaton oscillates around the minimum of its potential following the end of inflation. With the exception of the quadratic model, this oscillatory phase can trigger fascinating nonlinear phenomena whose properties depend on the detailed shape of the potential and the couplings of the inflaton to other fields. In particular, specific momentum modes of fields coupled to the inflaton and of the inflaton itself can undergo resonant amplification~\cite{Traschen1990,Kofman1994,Shtanov1994,Kofman1997}, fragmenting the initially homogeneous inflaton. 

Observations have ruled out purely monomial models and focus attention on potentials that are quadratic near their minima but grow more slowly at larger field values~\cite{Planck2018_inflation,BICEP:2021xfz}. This is the necessary condition for the existence of pseudo-solitonic field configurations known as \textit{oscillons}~\cite{Gleiser1993,Copeland1995,Amin2010,Amin2012,Lozanov2017} and  these form abundantly following resonance in many models where the minimum of the inflaton potential is surrounded by a sub-quadratic region. Oscillons are stable on scales longer than the post-inflationary Hubble time and the Universe can pass through an oscillon-dominated phase.  Moreover, resonance  (and oscillon formation, if it takes place) can source a high-frequency stochastic gravitational wave background~(SGWB)~\cite{Khlebnikov1997_grav,Easther2006_1,Easther2006_2,Easther2007,Garcia-Bellido2007_2,Dufaux2007,Dufaux2008,Zhou2013, Amin2018,Hiramatsu2020,Lozanov2022} and may even lead to the formation of primordial black holes~\cite{Cotner2018}.

In the absence of parametric resonance  the dominant inflaton interactions are purely gravitational and the amplitude of the inflaton  oscillations will slowly decrease. If the inflaton oscillates around a quadratic minimum a period of effectively matter-dominated expansion occurs and perturbations in the inflaton grow gravitationally on sub-horizon scales~\cite{Jedamzik2010_dens,Easther2010}. 
In this regime, the self-gravitating inflaton condensate can be described by the nonrelativistic Schrödinger-Poisson equations~\cite{Musoke2019}. Simulations have shown that growing inflaton fluctuations can collapse  into gravitationally bound structures prior to thermalization~\cite{Musoke2019,Niemeyer2019,Eggemeier2020,Eggemeier2021,Eggemeier2022}.  This can be accompanied by the generation of a potentially nontrivial SGWB~\cite{Eggemeier2022,Jedamzik2010_grav}.

There is a clear analogy between  nonlinear gravitational collapse in the post-inflationary universe and  the gravitational evolution of axion-like or fuzzy dark matter~(FDM)~\cite{Hu2000,Hui2017} in the late-time universe.\footnote{Also known as ultralight dark matter.} In particular, the same methods used to solve the Schrödinger-Poisson equations in FDM structure formation simulations can be exploited to study the formation of gravitationally bound inflaton structures in the early matter-dominated epoch. Naively, a numerical simulation must spatially resolve the de Broglie wavelength to reveal phenomena arising from wave interference such as gravitational solitonic objects known as boson stars that are of special interest in the context of axion-like dark matter; see e.g. Refs.~\cite{Tkachev1986,Tkachev1991,Schive2014_nature,Schive2014_prl,Schwabe2016,Veltmaat2018,Mocz2017,Mocz2018,Levkov2018,Pyultralight,Eggemeier2019PRD,Jiajun2020,Schwabe2020,Schwabe2021,Chan2022,Gosenca2023,Dmitriev2023}. Achieving this throughout a large-scale cosmological simulation is computationally infeasible. However, on large scales there is a correspondence between the coarse-grained Schrödinger-Poisson and the Vlasov-Poisson equations~\cite{Widrow1993,Uhlemann2014}, so only N-body simulations are needed if it is unnecessary to resolve the wave-like behaviour at small scales.   

Until now, post-inflationary gravitational collapse has been primarily  explored with a  quadratic inflaton potential. This suppresses  self-resonant field amplification and ensures that the inflaton condensate fragments gravitationally. Large N-body simulations confirm that  \textit{inflaton halos} form roughly 17 $e$-folds after the end of inflation  in this scenario, with typical masses of grams to kilograms~\cite{Eggemeier2020}.  Subsequent simulations with adaptive mesh refinement~(AMR) which solved the Schrödinger-Poisson equations directly in high density regions showed the formation of  dense solitonic cores, so called \textit{inflaton stars}, in the center of the inflaton halos~\cite{Eggemeier2021}. These simulations suggest that if the matter-dominated era lasts  long enough some inflaton stars might collapse to form primordial black holes~(PBHs), a possibility further explored in  Refs.~\cite{Padilla2022,Hidalgo2022}. 
Moreover, the  formation and subsequent interactions of inflaton structures  leads to gravitational wave emission. This spectrum was computed directly from simulations up to 23 $e$-folds after the end of inflation and  estimated semi-analytically for reheating temperatures as low as $100\,\mathrm{MeV}$~\cite{Eggemeier2022}. In extreme cases the resulting signal might be detectable by the Einstein Telescope~\cite{ET} or a space-based, BBO-like instrument~\cite{BBO}. 

That said, the  quadratic inflaton potential assumed in the  post-inflationary structure formation simulations of Refs.~\cite{Eggemeier2020,Eggemeier2021,Eggemeier2022} is excluded observationally~\cite{Planck2018_inflation}. Furthermore, the growth of density fluctuations proceeds slowly, leading to a long gap between the end of inflation and the beginning of gravitational collapse. For that reason, we wish to consider more realistic single-field inflationary models whose potentials grow less steeply away from the quadratic minimum. 
The primary inflationary perturbation spectrum will differ between these scenarios and a quadratic model but, more importantly, these  potentials typically support self-resonance and thus quickly generate significant  overdensities.  Hence, post-inflationary gravitational structure formation   can begin much earlier than  in the  quadratic case~\cite{Eggemeier2020,Eggemeier2021,Eggemeier2022} and the overall nonlinear matter-dominated phase can last much longer and is thus more likely to have detectable consequences. Therefore, simulating the full dynamics of the post-inflationary phase in realistic models is a critical challenge -- in addition to the need for a complete account of the dynamics of these observationally favoured scenarios, this novel phase can lead to specific predictions that may be experimentally testable.
 
In this work we present the first combined simulations that follow the inflaton field through parametric self-resonance into gravitational fragmentation. We focus on the single-field axion monodromy inflationary model~\cite{Silverstein2008,McAllister2008} (which may itself over-produce primordial tensor perturbations relative to the latest constraints~\cite{BICEP:2021xfz}) starting the simulations immediately after the end of inflation in a Klein-Gordon solver and running through to the end of resonance. At this point the inflaton field configuration is mapped into the nonrelativistic wave function and used to initialize the AMR-enabled  Schrödinger-Poisson solver from \textsc{AxioNyx}~\cite{Schwabe2021}. We observe the formation of inflaton halos with masses of up to $10^4$ Planck masses 7.6 $e$-folds after the end of inflation. Using a halo finder we extract the evolution of the halo mass function and confirm the formation of inflaton stars in the halo centers.  Compared to simulations with a  quadratic inflaton potential~\cite{Eggemeier2020,Eggemeier2021,Eggemeier2022}, collapsed structures form roughly 10 $e$-folds of expansion earlier in this scenario. This  will likely have a significant effect on the strength of gravitational wave emission and on potential PBH formation. 

The one key restriction on our simulations is that we have deliberately chosen model parameters such that oscillon formation does not follow resonance. In contrast to the gravitationally bound solitons, oscillons are supported by nonlinear self-interactions of the field itself, and although long-lived they do eventually decay relativistically.  These might be describable at the wave function level for oscillons with nonrelativistic frequencies by adding nonlinear interactions (at least during the period in which resonance is operating~\cite{Amin:2019ums}) to the Schrödinger-Poisson equation but this is not possible with generic oscillons (see e.g. Refs.~\cite{Amin:2010jq,Amin:2019ums,vanDissel:2023zva}). 

The structure of this paper is as follows. In \cref{sec:scenario} we describe the overall scenario with an era of gravitational structure formation following a period of parametric self-resonance and present our chosen model. The simulation setup of the resonance simulations and their results are discussed in \cref{sec:KGsims}. The use of the endpoint of these calculations to initialize  a Schrödinger-Poisson solver is  described in \cref{sec:SPsims}, along with the outcome of these simulations. We discuss our results and conclude in \cref{sec:conclusion}.

\section{Scenario}
\label{sec:scenario}

In single-field models of inflation a scalar  field $\varphi$ drives the accelerated expansion.  A homogeneous inflaton field with effective potential $V(\varphi)$ in a flat Friedmann-Lemaitre-Robertson-Walker space-time obeys the Klein-Gordon equation
\begin{align}
    \Ddot{\varphi} + 3H\dot\varphi + \frac{\mathrm{d}V}{\mathrm{d}\varphi} = 0\,,
    \label{eq:kg_eq}
\end{align}
while the expansion of space is described by the Friedmann equation
\begin{align}
    H^2 = \frac{1}{3M_\mathrm{Pl}^2}\left(\frac{1}{2}\dot\varphi^2 + V(\varphi)\right)\,.
\end{align}
As usual, $H=\dot a/a$ is the Hubble parameter, $a$ is the scale factor and $M_\mathrm{Pl} = (8\pi G)^{-1/2}$ is the reduced Planck mass. 
We consider a potential inspired by the axion monodromy model~\cite{Silverstein2008,McAllister2008,Amin2012} 
\begin{align}
    V(\varphi) = \frac{m^2M^2}{2\alpha}\left[\left(1+\frac{\varphi^2}{M^2}\right)^\alpha-1\right]\,,
    \label{eq:inflaton_potential}
\end{align}
where $m$ denotes the inflaton mass and $\alpha$ is treated as a free parameter, but must be less than unity for the model to match observations. For small $\varphi$ the potential approaches $V(\varphi)\simeq m^2\varphi^2/2$ while $V(\varphi)\sim m^2 M^{2-2\alpha}\varphi^{2\alpha}/(2\alpha)$ for large $\varphi$ and $M$ parametrizes the crossover scale between these two regimes. 
 
\begin{figure*}
    \centering
    \includegraphics[width=\textwidth]{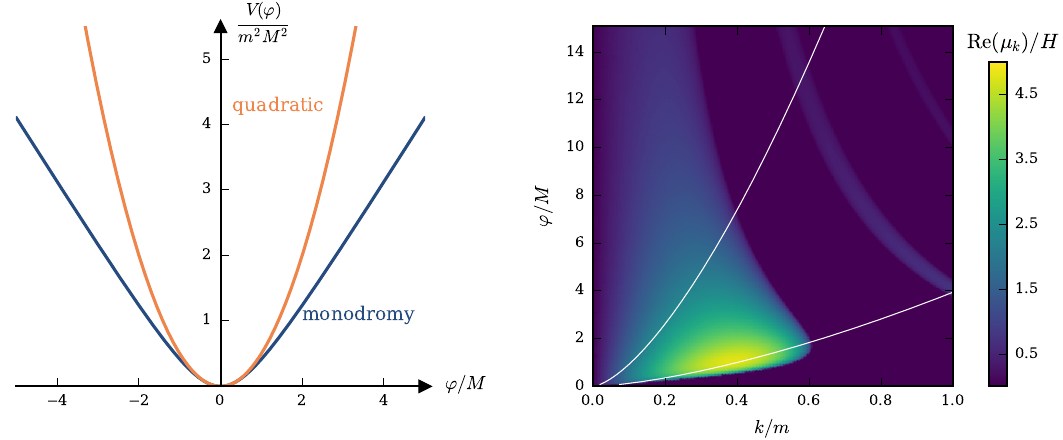}
    \caption{Left: Inflationary monodromy potential for parameters $\alpha=1/2$, $M=0.05M_\mathrm{Pl}$ that supports parametric self-resonance. For comparison, we show a  quadratic $m^2\varphi^2$ potential. Right: Floquet stability diagram for the monodromy potential shown on the left. The white lines visualize the evolution of two physical wave numbers $k/a$ in an expanding universe. They mark the boundaries of the broad resonance peak shown in \cref{fig:kg_ps_evolution}. }
    \label{fig:potential_floquet}
\end{figure*}

After inflation ends the inflaton oscillates around the minimum of \cref{eq:inflaton_potential}. When $\varphi\lesssim M$ the potential is approximately quadratic and the post-inflationary growth of the scale factor matches that of a matter-dominated universe. Provided the field is not disrupted by  resonance the scale factor grows as $a(t)\sim t^{2/3}$ while the field amplitude and the Hubble parameter decrease as $\varphi\sim a^{-3/2}$ and $H\sim a^{-3/2}$, respectively.  
Depending on the couplings between  the inflaton and other fields this early matter-dominated era can last for tens of $e$-folds. 
This leaves substantial room for parametric self-resonance and an extensive era of gravitational structure formation.

\subsection{Parametric self-resonance}
\label{sec:parametric_resonance}

In contrast to a  quadratic potential, an inflaton potential with the form of 
\cref{eq:inflaton_potential} supports parametric self-resonance. The strength of the parametric amplification depends on $\alpha$ and $M$~\cite{Amin2012}. To gain an understanding of the parametric growth of perturbations for the given potential one can make use of Floquet theory to compute the corresponding instability diagram.  

Decomposing the inflaton field $\varphi = \bar\varphi + \delta\varphi$ into a homogeneous background $\bar\varphi$ and a perturbation $\delta\varphi$, the equation of motion for the homogeneous background field takes the form of \cref{eq:kg_eq} while in Fourier space the perturbations obey to first order
\begin{align}
    \delta\Ddot{\varphi}_k + 3H \delta\dot\varphi_k + \left(\frac{k^2}{a^2} + V^{\prime\prime}(\bar\varphi)\right)\delta\varphi_k = 0\,.
    \label{eq:perturbation_eq}
\end{align}
The periodically evolving background inflaton field enters this equation and acts as a forcing term for the inflaton perturbations. In this case Floquet theory can be used to find solutions for $\delta\varphi_k$. 
In a non-expanding universe ($a=1$, $H=0$) the solutions are of the form~\cite{Amin2014_review,Amin2012}
\begin{align}
    \delta\varphi_k = P_+(t)e^{\mu_kt} + P_-(t)e^{-\mu_kt}\,,
    \label{eq:floquet_solution}
\end{align}
where $P_\pm$ are periodic functions and $\pm\mu_k$ are the so-called Floquet exponents. They are complex numbers and determine the parametric amplification of the inflaton perturbations for a given scale $k$. If $\mathrm{Re}(\mu_k) > 0$ they grow exponentially. In an expanding universe the effective wave number is $k/a$ and modes move in and out of resonance so that strong  fluctuation growth needs $\mathrm{Re}(\mu_k) > H$.

Adopting the algorithm of Ref.~\cite{Amin2014_review} to calculate the Floquet exponents, the instability diagram for the potential in \cref{eq:inflaton_potential} is shown on the right-hand side of \cref{fig:potential_floquet} for   $\alpha=0.5$ and $M = 0.05M_\mathrm{Pl}$. 
The Hubble parameter is computed from $H^2 = V(\varphi)/(3M_\mathrm{Pl}^2)$, where $\varphi$ is understood as the amplitude of the oscillating field in this expression. 
With a maximum of $\mathrm{Re}(\mu_k)/H = 4.8$, this model allows for moderate resonance while  the formation of oscillons is typically seen with $\mathrm{Re}(\mu_k)/H\gtrsim 10$~\cite{Amin2012,Amin2014_review}. With this choice we can  simulate the gravitational collapse of resonantly amplified inflaton perturbations with the nonrelativistic Schrödinger-Poisson equations.

\subsection{Gravitational structure formation}

If the reheating temperature is sufficiently low the inflaton condensate  eventually becomes gravitationally unstable. In principle, one could solve the full relativistic Einstein-Klein-Gordon equation but this would be  infeasible  over many $e$-folds. Instead, we make use of a nonrelativistic approximation that relies on integrating out the rapid oscillations of the inflaton. This choice means we cannot capture the dynamics of oscillons in most models but it greatly facilitates the analysis of scenarios in which they are absent. 

Since $H\sim a^{-3/2}$ in the post-inflationary epoch the relationship $m\gg H$ holds a few $e$-folds after the end of inflation. In this limit one can use the WKB approximation to express the inflaton in terms of a slowly varying, complex field $\psi$~\cite{Widrow1993},
\begin{align}
    \varphi = \frac{\hbar}{\sqrt{2}m a^{3/2}}\left(\psi e^{-imt/\hbar} + \psi^\ast e^{imt/\hbar}\right)\,.
    \label{eq:wkb_approx}
\end{align}
This ansatz factors out the fast oscillations of $\varphi$ using the approximation  $|\dot\psi| \ll m |\psi|$.  The wave number of small perturbations is effectively an oscillation time and the nonrelativistic limit is valid when the  wave number $k$ satisfies  $k/a\ll m$ on scales of interest. Finally, the field values must be small enough so that  \cref{eq:inflaton_potential} can be approximated as $V(\varphi)\simeq m^2\varphi^2/2$.  Under these conditions the evolution of the inflaton field under its own self-gravity is described by the nonrelativistic (comoving) Schrödinger-Poisson equations~\cite{Ruffini1969,Nambu1990QuantumPerturbations,Musoke2019}
\begin{align}
     i\hbar \partial_t\psi &= -\frac{\hbar^2}{2ma^2}\nabla^2\psi +  m \psi V_N\,, \label{eq:Schroedinger_eq}\\
     \nabla^2V_N &= \frac{4\pi G}{a}\left(\rho - \bar\rho\right)\,,
     \label{eq:poisson_eq}
\end{align}
where $\rho = |\psi|^2$ is the matter density of the inflaton field with mean $\bar\rho$. 

Moderate inflaton overdensities produced during self-resonance will grow and collapse into gravitationally bound inflaton halos and inflaton stars. Eventually, we expect that the gravitational fragmentation will break down when the Hubble parameter becomes comparable to the decay rate $\Gamma$ of the inflaton and the field decays into radiation. This reheating temperature is given by~\cite{Kofman1997}
\begin{align}
    T_\mathrm{rh} \simeq 0.55\left(\frac{100}{g_\ast}\right)^{1/4} \left(\Gamma M_\mathrm{Pl}\right)^{1/2}\,,
    \label{eq:reheating_temp}
\end{align}
where $g_\ast$ denotes the number of relativistic degrees of freedom.

\subsection{Parameter choice} 

Using the monodromy potential from \cref{eq:inflaton_potential} we evolve the inflaton field through a phase of parametric self-resonance by solving the Klein-Gordon equation. The final configuration then initializes the nonrelativistic wave function that is further evolved with the Schrödinger-Poisson equations. As noted, we choose the free parameters in the monodromy potential to avoid oscillon formation in the self-resonant phase so that the nonrelativistic approximation will hold. 

The observed amplitude of the  primordial curvature perturbations sets one  parameter in the monodromy potential. 
During slow-roll inflation, the potential in \cref{eq:inflaton_potential} can be approximated in the large-field limit $\varphi\gg M$.  
In this case, the  amplitude of the dimensionless curvature power spectrum at the end of inflation is~\cite{Amin2012}
\begin{align}
    \Delta^2_\mathcal{R} = \frac{(4\alpha N_\ast + 2\alpha)^{1+\alpha}}{96\pi^2\alpha^3}\left(\frac{m}{M_\mathrm{Pl}}\right)^2\left(\frac{M}{M_\mathrm{Pl}}\right)^{2-2\alpha}\,,
    \label{eq:monodromy_ps}
\end{align}
where $N_\ast$ denotes the number of $e$-folds astrophysical perturbations of interest left the horizon before the end of inflation. We will fix $N_\ast=55$  but  the precise value depends on the details of the post-inflationary evolution~\cite{Adshead2011}. 

For the sake of definiteness we set $\alpha=1/2$ which recovers the linear axion monodromy potential~\cite{Silverstein2008,McAllister2008}; the value of $M$ then controls the efficiency of resonance. Oscillon formation occurs after strong resonance, which requires $M\lesssim 0.01M_\mathrm{Pl}$~\cite{Amin2012,Lozanov2022}. However, significant resonance still occurs with $M=0.05M_\mathrm{Pl}$, as seen from the Floquet diagram in \cref{fig:potential_floquet} and we work with this value in what follows. Given that $\Delta_\mathcal{R} = 2.1\times 10^{-9}$~\cite{Planck2018_inflation}, we fix $m =  7.0 \times 10^{-5} M_\mathrm{Pl}$.

\section{Klein-Gordon simulations}
\label{sec:KGsims}

\begin{figure}
    \centering
    \includegraphics[width=\columnwidth]{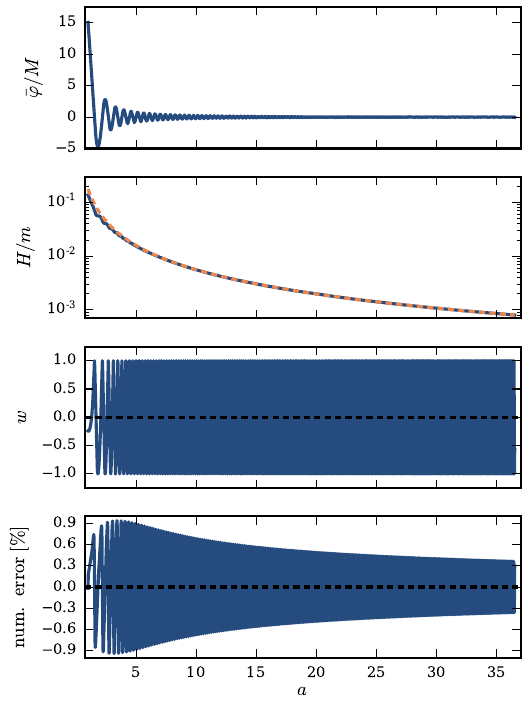}
    \caption{Evolution of the inflaton field $\bar\varphi$, the Hubble parameter $H$, the equation of state parameter $w$ and the numerical error ($H^2/(8\pi G\bar\rho/3)-1$) as a function of scale factor $a$. The dashed orange curve in the second upper panel corresponds to $H\sim a^{-3/2}$. } 
    \label{fig:kg_field_evolution}
\end{figure}

We implemented a Klein-Gordon finite-difference solver similar to \textsc{LatticeEasy}~\cite{Latticeeasy} in the \textsc{AMReX}~\cite{Amrex} framework.  This same package underlies the Schrödinger-Poisson solver of \textsc{AxioNyx}~\cite{Schwabe2021}, so we can follow the full evolution within a single codebase. However, in contrast to \textsc{AxioNyx}, the Klein-Gordon solver self-consistently solves the coupled Klein-Gordon-Friedmann equations, rather than using a fixed equation of state, which is necessary for consistency early in the simulation (see \cref{fig:kg_field_evolution}). 
The transition between the two solvers is made when the self-resonant amplification of resolved modes becomes negligible and the scales of interest are nonrelativistic.  For our specific scenario  this happens after $3.6$ $e$-folds of post-inflationary expansion.

\subsection{Simulation setup}

Following \textsc{LatticeEasy} we define computationally convenient  program variables for $\varphi$, length and time  as $\varphi_\mathrm{pr} = Aa^r\varphi_\mathrm{phys}$, $l_\mathrm{pr} = B l_\mathrm{phys}$ and $t_\mathrm{pr} = Ba^st_\mathrm{phys}$, respectively. The choice of scaling parameters $A$, $B$, $r$ and $s$ depends on the dominant term of the inflaton potential. Expressing it as $V = \lambda/\beta\varphi^\beta$, the scaling parameters are $A = \varphi_0^{-1}$, $B = \lambda^{1/2}\varphi_0^{-1+\beta/2}$, $r = 6/(2+\beta)$ and $s=3(2-\beta)/(2+\beta)$ where $\varphi_0$ is the initial value of the field $\varphi$~\cite{Latticeeasy}. 
For the potential in \cref{eq:inflaton_potential} with $\alpha=1/2$ it is  $B = m(M/\varphi_0)^{1/2}$, $r = 2$ and $s=1$.  
We consider the value of the homogeneous inflaton field at the end of slow-roll inflation as our choice for $\varphi_0$ and compute it numerically from setting the slow-roll parameter $\varepsilon(\varphi_0)=1$ which yields $\varphi_0 = 0.75\,M_\mathrm{Pl}$.  

\begin{figure}[t]
    \centering
    \includegraphics[width=\columnwidth]{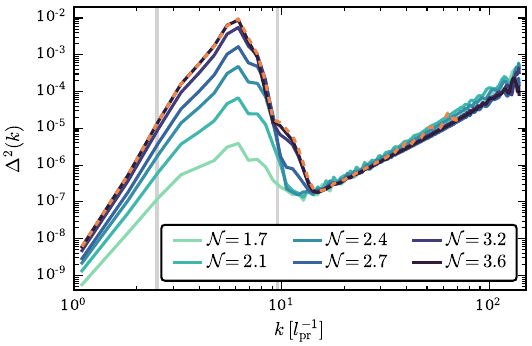}
    \caption{Evolution of the dimensionless density power spectrum from $\mathcal{N}=1.7$ to $3.6$ $e$-folds after the end of inflation. For comparison, the orange dashed curve shows the spectrum at $\mathcal{N}=3.6$ of a simulation with a $128^3$ lattice.   The two vertical gray lines mark the boundaries of the resonance peak whose evolution is shown in the Floquet diagram of \cref{fig:potential_floquet}. }
    \label{fig:kg_ps_evolution}
\end{figure}

\begin{figure*}
    \centering
    \includegraphics[width=\textwidth]{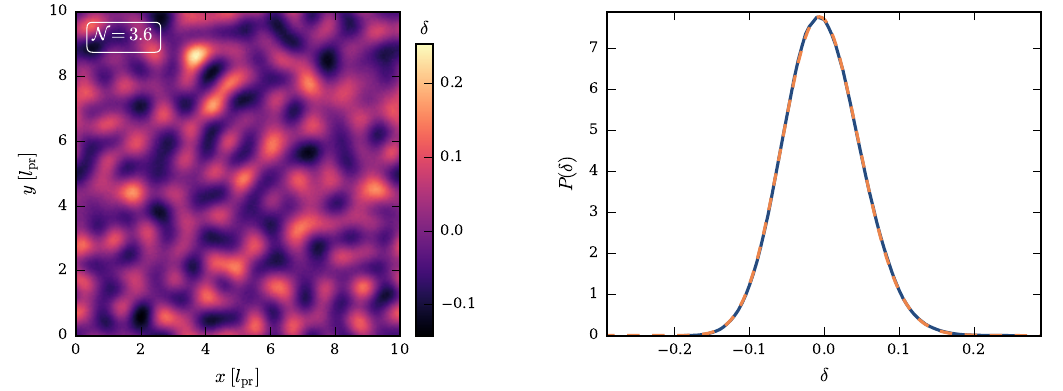}
    \caption{Left: slice through the center of the full simulation box showing the overdensity field $\delta$ at $\mathcal{N}=3.6$. Right: normalized distribution of the overdensity field (blue curve) at the same time. The dashed orange curve shows a log-normal distribution (see \cref{eq:lognormal_dist}) with parameters $\sigma_\delta=5.2\times 10^{-2}$ and $\mu_\delta=-4.1\times 10^{-3}$.}
    \label{fig:kg_final_density}
\end{figure*} 

We use a comoving box with sides of  length of $L=10\,l_\mathrm{pr}$ and solve the equations of motion on a $256^3$ grid.\footnote{Choosing a highly-resolved grid is not necessary as resonance occurs only on comparatively large scales for our specific scenario (see \cref{fig:kg_ps_evolution}).} The scale factor is normalized to $a(t_0)=1$ at the beginning of the simulations and evolved by solving the Friedmann equations. In contrast to the  fixed time step used in \textsc{LatticeEasy}, we set 
\begin{align}
    \Delta t = \frac{1}{2}\frac{\Delta x}{\sqrt{3a}}\,,
    \label{eq:timestep}
\end{align}
where $\Delta x$ is the cell width in physical coordinates. This choice guarantees that the Courant stability condition $\Delta t < \Delta x/\sqrt{3}$ is always fulfilled and  $\Delta t \sim a^{1/2}$ in comoving coordinates. Consequently, we have small time steps in the early stages of the simulation when self-resonance is active and larger steps when the energy density changes more slowly. This choice reduces the computational costs substantially compared to a fixed time step.   

We initialize the inflaton field in agreement with \textsc{LatticeEasy} and set $\mathrm{d}\bar\varphi_\mathrm{pr}/\mathrm{d}t_\mathrm{pr} = 0$.
The initial conditions are defined in momentum space and given by $\langle |\varphi_k|^2\rangle = 1/(2\omega_k)$ where $\omega_k^2 = k^2 + m^2$.  
The amplitude of each mode is fixed from a Gaussian random distribution; the phase of each mode is drawn randomly from a uniform distribution. The initial field values are then given by the inverse Fourier transform.
We iterate the initial conditions once, re-setting $\mathrm{d}a/\mathrm{d}t_\mathrm{pr}$ to ensure $H^2/(8\pi G \bar{\rho}/3) = 1$ initially. The subsequent violation of this condition furnishes a measure of the numerical error of the simulation, as discussed below.

\subsection{Self-resonant evolution}
 
\cref{fig:kg_field_evolution}  shows the evolution of $\varphi$, $H$ and the equation of state parameter $w$ during resonance. As expected, these quantities evolve as in a  matter-dominated universe with $\varphi\sim a^{-3/2}$, $H\sim a^{-3/2}$ and $\bar w = 0$. Furthermore, we estimate the energy conservation and thus the numerical error by computing the deviation between the ratio of $H^2$ and $8\pi G\bar\rho/3$ and unity. The numerical error is well under control over the entire simulation with a final deviation of $\sim 0.3\%$ from perfect energy conservation. 

The energy density at each grid point consists of a kinetic term, a gradient term and the field potential,
\begin{align}
    \rho = \frac{1}{2}\dot\varphi^2 + \frac{1}{2a^2}|\nabla\varphi|^2 + V(\varphi)\,.
    \label{eq:kg_energydensity}
\end{align}
Given the overdensity field $\delta = \rho/\bar\rho - 1$ we define the density power spectrum as $P(k) = V\langle|\delta_k|^2\rangle$ where $V=L^3$. For the purpose of studying the formation of gravitationally bound  structures in the post-resonance universe it is suitable to work with the dimensionless power spectrum
\begin{align}
    \Delta^2(k) = \frac{k^3}{2\pi^2}P(k)\,,
    \label{eq:dimless_powerspectrum}
\end{align}
as the threshold $\Delta^2(k) = 1$ indicates when a density perturbation on a given scale $k$ becomes nonlinear.

The evolution of $\Delta^2(k)$ during resonance is shown in \cref{fig:kg_ps_evolution} for $\mathcal{N}$ between $1.6$ and $3.6$ $e$-folds after the end of inflation. 
At $\mathcal{N}=1.7$ we observe the formation of a broad peak around $\sim 6\,l_\mathrm{pr}^{-1}$.
The amplitude of the resonant peak continues to grow until $\mathcal{N}=3.2$. Afterwards there are only minor changes to the shape of the power spectrum. This is consistent with the Floquet indices shown in \cref{fig:potential_floquet} which predict that the strongest growth occurs for comoving wave numbers within the resonance peak when the oscillating field  amplitude satisfies of $0.05\leq \bar\varphi/M\leq 0.8$. Note that the  strong increase of $\Delta^2\sim k^3$ at small scales that are not resonantly amplified is numerical shot noise and thus unphysical. To verify that it does not affect the physical outcome we performed a comparison run with the same initial conditions and a coarser $128^3$ grid. The corresponding power spectrum at $\mathcal{N}=3.6$ is shown by  the orange dashed curve  in \cref{fig:kg_ps_evolution} and it perfectly agrees with the final spectrum of the higher-resolution run. 

After $\mathcal{N}=3.6$ $e$-folds of expansion $H/m \sim 10^{-3}$  (see \cref{fig:kg_field_evolution}) and the highest-momentum modes satisfy $k/m < 0.1$. This permits the use of the nonrelativistic approximation and we make the transition to the Schrödinger-Poisson description at this time.

A slice through the overdensity field $\delta$ at $\mathcal{N}=3.6$ is visible in the left panel of \cref{fig:kg_final_density} showing inhomogeneities of $\mathcal{O}(10^{-1})$. The distribution of the overdensity field is shown in the right panel of \cref{fig:kg_final_density}, and this is in good agreement with the log-normal distribution
\begin{align}
    P(\delta) = \frac{1}{\delta\sigma_\delta\sqrt{2\pi}}\exp\left(-\frac{(\ln\delta - \mu_\delta)^2}{2\sigma_\delta^2}\right)
    \label{eq:lognormal_dist}
\end{align}
with $\sigma_\delta=5.2\times 10^{-2}$ and $\mu_\delta=-4.1\times 10^{-3}$.

\section{Schrödinger-Poisson simulations}
\label{sec:SPsims}

Having evolved the inflaton field through a resonant phase we initialize the Schrödinger-Poisson solver and evolve the universe through an additional 4 $e$-folds of expansion.  Numerical solutions of Schrödinger-Poisson equations must ensure that the de Broglie wavelength is properly resolved. We make use of AMR techniques to ensure that the resolution requirement is fulfilled at all times. 

\subsection{Simulation setup}

\begin{figure*}
    \centering
    \includegraphics[width=\textwidth]{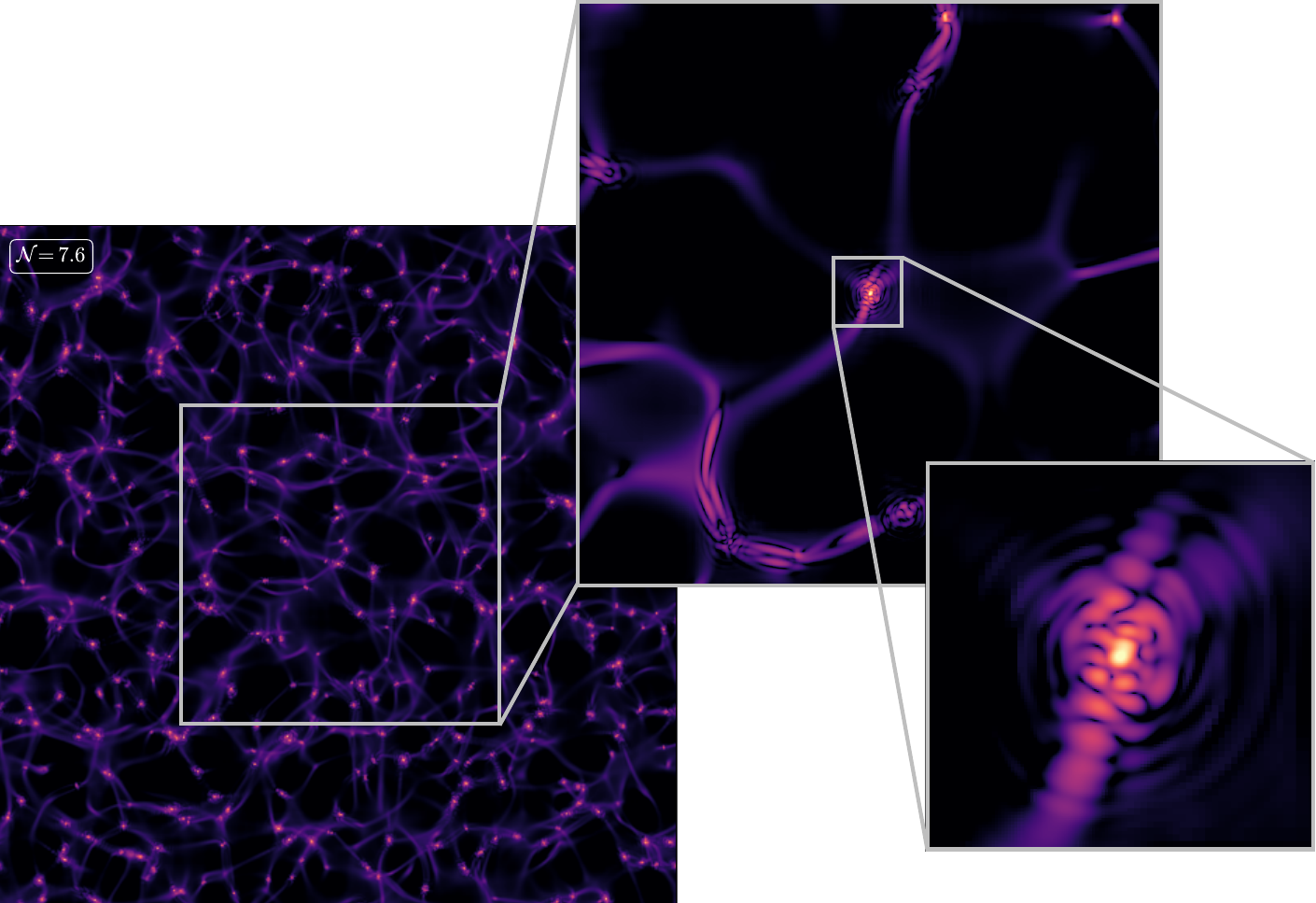}
    \caption{Visualization of the inflaton field $\mathcal{N}=7.6$ $e$-folds after the end of inflation from larger to smaller scales. On the left, the projected inflaton density of the full simulation box ($L = 5.53\times 10^9\,l_\mathrm{Pl}$) centered on a selected halo is shown. A slice through the density maximum of this halo covering the area of the grey square ($L=2.76\times 10^9\,l_\mathrm{Pl}$) illustrates the occurrence of interference patterns in the filaments and the typical granular structure inside the halos. In the enlargement of the selected halo on the right ($L=1.63\times 10^8\,l_\mathrm{Pl}$) one can identify the prominent solitonic core in its center.}
    \label{fig:SP_density_projection}
\end{figure*}

Using \cref{eq:wkb_approx} and dropping small terms\footnote{Note that this requires a choice of convention to remove the redundancy in inverting \cref{eq:wkb_approx}; see Ref.~\cite{Salehian2020} for details.} we initialize the wave function $\psi$ from the inflaton field $\varphi$ and its time derivative $\dot\varphi$ as 
\begin{align}
    \psi = \frac{ma^{3/2}}{\sqrt{2}\hbar}\left(\varphi + i\frac{\hbar}{m}\dot\varphi\right)e^{imt/\hbar}\,.
\end{align}
To perform cosmological simulations \textsc{AxioNyx} takes the value of the Hubble parameter $H_\mathrm{end}$ at the end of the simulation as an input parameter, at which point  the  scale factor is normalized to unity whereas the Klein-Gordon solver works internally with a scale factor which satisfies $a=1$ at the end of inflation. Since we consider four $e$-folds of additional expansion, we set $a_\mathrm{init}/a_\mathrm{end} = e^{-4}$ in \textsc{AxioNyx}. The Hubble parameter is thus given by its value $H_\mathrm{init}$ at the end of the Klein-Gordon run via $H_\mathrm{end} = H_\mathrm{init}(a_\mathrm{init}/a_\mathrm{end})^{3/2} = H_\mathrm{init}e^{-6}$. 

The comoving size of the simulation box  ($L=10\,l_\mathrm{pr}$) remains unchanged. Given the total expansion of $7.6$ $e$-folds, we define the box size at $\mathcal{N}=7.6$ in physical coordinates as the length unit for the \textsc{AxioNyx} run, so $l_u = 5.53\times 10^9\,l_\mathrm{Pl}$ where $l_\mathrm{Pl}$ is the Planck length. 
Choosing the mass unit as $m_u = 2.3\times 10^{-2}\,M_\mathrm{Pl}$ and taking the gravitational constant to be $G = 10^{-10}\,l_u^3/(m_u t_u^2)$, the time unit is $t_u = 6.06\times 10^{10}\,t_\mathrm{Pl}$ where $t_\mathrm{Pl}$ denotes the Planck time.
The Hubble parameter and the mean matter density $\mathcal{N}=7.6$ $e$-folds after the end of inflation are $H_\mathrm{end} = 1.69\,t_u^{-1}$ and $\bar\rho_\mathrm{end} = 3.42\times 10^9\,m_u/l_u^3$. 

A fourth-order Runge-Kutta finite-difference algorithm is used to solve the Schrödinger-Poisson equations on the $256^3$ root grid and on the refined levels. We allow four levels of refinement with a refinement factor of two at each step. Refinement occurs when the comoving density of a cell exceeds the  thresholds $1.5\bar\rho_\mathrm{end}$ on the root grid, $5\bar\rho_\mathrm{end}$ on the first level, $10\bar\rho_\mathrm{end}$ on the second level and $20\bar\rho_\mathrm{end}$ on the third level. 
This choice guarantees that the de Broglie wavelength is always resolved.
At the end of the simulation $60.2\%$ of the  volume is refined on the first level, while the second, third and fourth levels cover $15.2\%$, $4.1\%$ and $1.1\%$ of the  domain, respectively.

\subsection{Gravitational evolution}

As expected, gravitationally bound inflaton halos form after the phase of self-resonance. 
Resonance ends with typical overdensities smaller than $\mathcal{O}(10^{-1})$ and so the first structures start to collapse gravitationally after three $e$-folds of expansion.
For $\mathcal{N}\gtrsim 6.9$ an increasing number of halos then form, with masses of up to \mbox{$6\times 10^4\,M_\mathrm{Pl}$}.  A visualization of the full simulation volume that exhibits the rich, highly nonlinear structure is shown in \cref{fig:SP_density_projection}. Overdensities of $\mathcal{O}(10^5)$ are reached in this scenario.

\begin{figure}[t]
    \centering
    \includegraphics[width=\columnwidth]{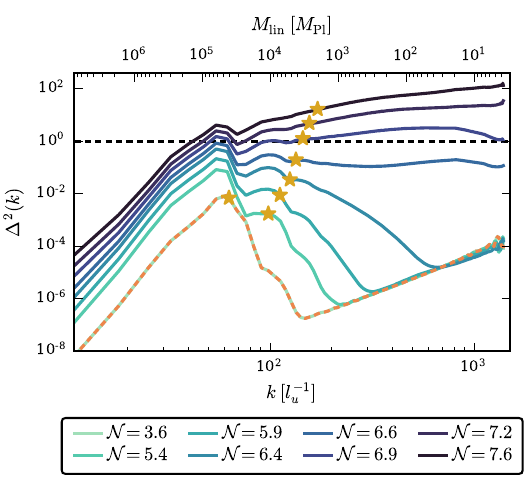}
    \caption{Evolution of the dimensionless density power spectrum from $\mathcal{N}=3.6$ to $7.6$ $e$-folds after the end of inflation. The dashed orange curve corresponds to the final power spectrum of the Klein-Gordon run. The stars illustrate the location of the comoving Jeans scale $k_J$ (see \cref{eq:jeans_scale}) at the respective times. The upper axis shows the characteristic linear mass for each wave number (see text for details).}
    \label{fig:SP_ps_evolution}
\end{figure}

The evolution of the dimensionless density power spectrum is shown in \cref{fig:SP_ps_evolution}. As a consistency check for the conversion of the inflaton field into the wave function, we compare the power spectrum computed from the density $\rho = |\psi|^2$ at initialization (light green curve) with the final power spectrum of the Klein-Gordon run (dashed orange curve) and they agree well with each other. 

Initially, the comoving Jeans scale 
\begin{align}
    k_J = \left(16\pi G a\bar\rho\frac{m^2}{\hbar^2}\right)^{1/4}
    \label{eq:jeans_scale}
\end{align}
coincides with the position of the resonance peak in the power spectrum.\footnote{We understand this to be a coincidence, but it is possible further study could reveal a deeper connection between the two scales.} Linear gravitational growth only takes place on scales $k < k_J$ while density perturbations oscillate on scales $k > k_J$.\footnote{Since the maximum amplitude of the dimensionless power spectrum is significantly below $\Delta^2 = 1$ linear perturbation theory is applicable.} Thus, the gravitational growth of structures on scales smaller than the resonance peak is initially suppressed. However, with growing $\mathcal{N}$ the comoving Jeans scale slowly increases and gravity starts to dominate on increasingly smaller scales, eventually leading to the collapse of the respective density fluctuations. 
It is possible to assign a characteristic linear mass $M_\mathrm{lin}$ to each wave number $k$ by considering a sphere with radius $r_\mathrm{lin} = \lambda/2 = \pi/k$  enclosing a region with the average density $\bar\rho$, so  $M_\mathrm{lin} = 4\pi/3 \bar\rho r_\mathrm{lin}^3$. This mass provides an estimate for the actual halo mass when a density fluctuation on a given scale becomes nonlinear at $\Delta^2\simeq 1$ and it is included in \cref{fig:SP_ps_evolution} as a secondary axis.

During the first three $e$-folds of post-resonant expansion we observe an overall increase in power on scales $k < k_J$ due to the gravitational growth of density perturbations. Additionally, a prominent peak develops at $k\sim 10^2\,l_u^{-1}$ slightly below the Jeans scale. Concurrently, there is a strong enhancement in the power spectrum on scales $k > k_J$ that cannot be explained by gravitational effects.\footnote{Note that the numerical shot-noise which originates from the Klein-Gordon simulation cannot be the origin of the observed growth of power on small scales. If that were the case, one would expect the transfer of power from the smallest to larger scales.  However, the shape of the spectrum on the smallest resolved scales does not evolve until $\mathcal{N}=6.4$ while a power increase can be observed for larger scales.}
Instead, this excess power originates from \textit{nonlinear} processes of wave interference on scales of the de Broglie wavelength (see interference patterns along filaments and in halos in \cref{fig:SP_density_projection}) that is also present in the power spectrum evolution of FDM simulations~\cite{Mocz2019,May2021,May2022}. 
Its occurrence on smaller and smaller scales with increasing $\mathcal{N}$ can be explained phenomenologically via the growth of structures on larger scales which results in an increasing local velocity $v$ and thus a decreasing $\lambda_\mathrm{dB} = 2\pi\hbar/(mv)$. 
Hence, shell-crossing events and eventually the virialization of collapsed structures produce a particularly strong increase of power on scales above the Jeans scale.

At $\mathcal{N}=6.6$ the amplitude of the resonance peak at $k\sim 60\,l_u^{-1}$ reaches the threshold $\Delta^2=1$. Accordingly, we observe the formation of the first gravitationally bound inflaton halos shortly afterwards at $\mathcal{N}=6.9$ by which time the aforementioned second peak located slightly above the Jeans scale has also become nonlinear. The mass of the just collapsed objects can be estimated to be $M_\mathrm{lin}\sim 5\times 10^4\,M_\mathrm{Pl}$. When $\mathcal{N}=7.6$  an increasing number of modes have become nonlinear and one  expects halo masses ranging from $M_\mathrm{lin}\sim 3\times 10^3\,M_\mathrm{Pl}$ to $M_\mathrm{lin}\sim 2\times 10^5\,M_\mathrm{Pl}$.

\begin{figure}
    \centering
    \includegraphics[width=\columnwidth]{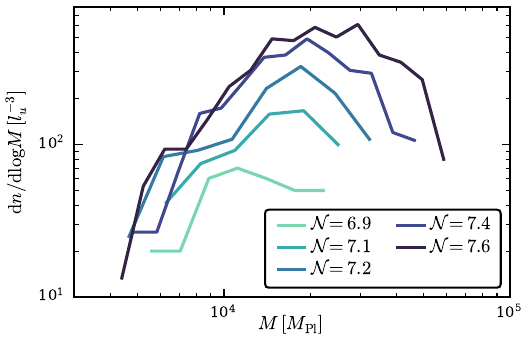}
    \caption{Evolution of the inflaton halo mass function from $\mathcal{N}=6.9$ to $7.6$ $e$-folds after the end of inflation.}
    \label{fig:hmf_evolution}
\end{figure}

We are not aware of any publicly available halo finder that operates on a discretized density field. However, we can sample the density field with $N=256^3$ particles by placing one particle per cell with a mass that corresponds to the cell density, located at a random position within the cell. We then use the \textsc{Hop} halo finder~\cite{HOP} to locate the inflaton halos in the simulation volume and  determine their masses.  We performed this sampling at different simulation snapshots and  verified that the power spectra computed from the particle mass density field agrees with the original spectra at the corresponding times.\footnote{We found general agreement between the halo mass function obtained from our early-universe N-body simulations~\cite{Eggemeier2020} and the mass function computed from the corresponding density field, validating the halo finding approach used here.}
The minimum particle number in a halo is set to 20. By computing the enclosed density as the radius is increased the virial radius $r_\mathrm{vir}$ of a halo is obtained when the enclosed density matches $\rho_\mathrm{vir} = \Delta_\mathrm{vir}\bar\rho$, where $\Delta_\mathrm{vir}=18\pi^2$ for a matter-dominated background and $\bar\rho$ is the mean matter density. The virial mass of the halos is then given by $M_\mathrm{vir} = 4\pi/3\Delta_\mathrm{vir}\bar\rho r_\mathrm{vir}^3$.  

The inflaton halo mass function (HMF), the comoving number density of halos per logarithmic mass interval, is shown in \cref{fig:hmf_evolution} for several values of $\mathcal{N}$. 
At $\mathcal{N}=6.9$, right after the first formation of halos, the HMF contains halos with masses of up to $M_h=2\times 10^4\,M_\mathrm{Pl}$ and is peaked at $\sim 10^4\,M_\mathrm{Pl}$. These halos presumably originate from the collapse of nonlinear density fluctuations that correspond to the broad resonance peak (see \cref{fig:SP_ps_evolution}). 
With increasing $\mathcal{N}$, more and more inflaton halos are produced extending the covered mass range to both lower and higher masses. At the final snapshot, we observe halo masses from \mbox{$M_h\sim 4\times 10^3\,M_\mathrm{Pl}$} to $M_h\sim 6\times 10^4\,M_\mathrm{Pl}$ and the HMF peaks  at $M_h\sim 2\times 10^4\,M_\mathrm{Pl}$. 

\begin{figure*}
    \centering
    \includegraphics[width=\textwidth]{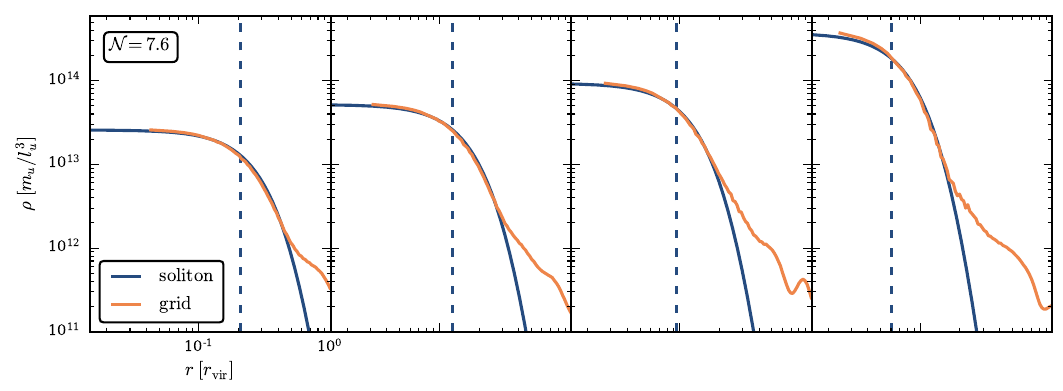}
    \caption{Radial density profiles of four inflaton halos $\mathcal{N}=7.6$ $e$-folds after the end of inflation with masses (from left to right) $M_h\in [1.1,\,2.8,\,4.3,\,5.8]\times 10^4 \,M_\mathrm{Pl}$. The orange curves display the simulation data, the blue curves correspond to the theoretical soliton density profiles from \cref{eq:soliton_dens_profile} and the blue vertical dashed lines mark the core radius $r_\ast$. The corresponding core masses from \cref{eq:soliton_mass} are (from left to right) $M_\ast\in [2.5,\,3.0,\,3.5,\, 5.0]\times 10^3\,M_\mathrm{Pl}$.}
    \label{fig:densprofiles}
\end{figure*}

Note that the halo masses do not precisely match the estimated masses $M_\mathrm{lin}$ from the analysis of the dimensionless density power spectrum. 
For example, \mbox{$M_\mathrm{lin} \sim 5\times 10^5\,M_\mathrm{Pl}$} for collapsed scales of the broad resonance peak at $\mathcal{N}=6.9$ while the maximum halo mass is $M_h\sim 2\times 10^4\,M_\mathrm{Pl}$ at that time. Nevertheless, the evolution of the HMF is understandable from the evolution of the power spectrum as it confirms the expectation that an initially narrow distribution of halo masses will extend to both lower and higher masses. Consequently, one can assume that both lower- and higher-mass halos will continue to form when $\mathcal{N}>7.6$. 
This is in contrast to non-resonant post-inflationary structure formation~\cite{Musoke2019,Eggemeier2020,Eggemeier2021,Eggemeier2022} where higher-mass objects emerge after lower-mass halos via bottom-up structure formation.

Beyond the halos themselves, our high-resolution simulations reveal the
formation of solitonic cores in the center of the halos. These inflaton stars are surrounded by incoherent granular density fluctuations associated with the wave function, as can be seen in \cref{fig:SP_density_projection}. Given the analogy with FDM, the expected radial density profile of an inflaton star is~\cite{Schive2014_nature,Schive2014_prl}
\begin{align}
    \rho_\ast(r)\simeq \rho_{\ast,0}\left(1+0.091\left(\frac{r}{r_\ast}\right)^2\right)^{-8}\,,
\end{align}
where $r_\ast$ denotes the physical core radius. This is defined to be the point at which the density is half of the central density,  
\begin{align}
    \rho_{\ast,0} = 4.6\times 10^{13}\left(\frac{7.0\times 10^{-5}M_\mathrm{Pl}}{m}\right)^2\left(\frac{10^{-3}\,l_u}{r_\ast}\right)^4\frac{m_u}{l_u^3}\,.
    \label{eq:soliton_dens_profile}
\end{align}
Defining the mass $M_\ast$ of an inflaton star as the mass inside $r_\ast$, we have 
\begin{align}
    M_\ast = 1.27\times 10^5\left(\frac{7.0\times 10^{-5}M_\mathrm{Pl}}{m}\right)^2\left(\frac{10^{-3}\,l_u}{r_\ast}\right)\,m_u\,.
    \label{eq:soliton_mass}
\end{align}
The radial density profiles of four different halos with masses ranging from $M_h = 1.1\times 10^4\,M_\mathrm{Pl}$ to \mbox{$M_h = 5.8\times 10^4\,M_\mathrm{Pl}$} at $\mathcal{N}=7.6$  are shown in \cref{fig:densprofiles}. 
The inner profiles are in good agreement with  \cref{eq:soliton_dens_profile}, confirming the existence of inflaton stars in the post-resonance, matter-dominated universe.  Via \cref{eq:soliton_mass}, the core masses for the density profiles shown in \cref{fig:densprofiles} range from $M_\ast = 2.5\times 10^3\,M_\mathrm{Pl}$ to $M_\ast = 5.0\times 10^3\,M_\mathrm{Pl}$. 

Given the virial velocity $v_\mathrm{vir}$ of their host halos we can infer the core masses via $M_\ast = \hbar v_\mathrm{vir}/(mG)$ which is also known as the core-halo mass relation. The inflaton star masses obtained from \cref{eq:soliton_mass} are 5-26\% lower than the prediction from the core-halo relation that is expected to be valid when the core is in virial equilibrium with its host halo. However, the corresponding power-law relationship $M_\ast\sim M_h^{1/3}$ appears to be set-up dependent~\cite{Zagorac2022,Kendall2023} and need not apply exactly to this case.

\section{Conclusions and discussion}
\label{sec:conclusion}

We have performed first-of-their-kind combined simulations that evolve the inflaton field through a phase of parametric self-resonance into the era of gravitational  structure formation. This was achieved by transitioning between the Klein-Gordon and Schrödinger-Poisson equations following the end of resonance, providing a computationally efficient strategy that pivots between solving the nonlinear field dynamics and the gravitational physics. 

In this initial investigation, we considered a specific inflationary model with no couplings to other fields. The free parameters in the  potential were selected to allow for a significant period of resonance while suppressing oscillon formation, as these metastable field configurations cannot be evolved in the  Schrödinger-Poisson system which does not include any non-gravitational self-interactions of the field. The simulations were initialized directly after the end of inflation and evolved for in total $7.6$ $e$-folds of expansion. 

During  self-resonance inflaton field fluctuations with specific wavelengths are amplified as they pass through the resonance bands, leading to a highly scale-dependent spectrum of overdensities. The growth of the dimensionless density power spectrum ceases $\mathcal{N}\approx 3.2$ $e$-folds after the end of inflation,  at which point resonance effects are thus negligible. We stopped the Klein-Gordon simulation at $\mathcal{N}=3.6$ and checked that the resolved scales are nonrelativistic and the Schrödinger-Poisson description is valid. 

The inflaton field was then mapped into a nonrelativistic wave function and the  Schrödinger-Poisson equations were solved to study the gravitational fragmentation of the inflaton field in the post-resonance epoch. 
Using AMR  with four levels of refinement we spatially resolved the formation of inflaton structures throughout the full simulation volume. The first inflaton halos form after roughly three $e$-folds of additional expansion, i.e. at $\mathcal{N}=6.9$. In contrast to typical structure formation simulations where lower-mass halos are usually produced earlier than higher-mass objects, low- and high-mass inflaton halos in our simulation can emerge at similar times. This is a consequence of  the broad  peak in the density power spectrum generated by resonance, which means that intuitions from structure formation scenarios with a near scale-free power spectrum need not be reliable.
Our simulation ended at $\mathcal{N}=7.6$ and at this moment the inflaton halos cover masses up to $6\times 10^4\,M_\mathrm{Pl}$. In addition, our high-resolution simulations showed central soliton-like configurations at the centers of these halos, confirming the rapid formation of inflaton stars in these halos.

Gravitationally bound structures form roughly 10 $e$-folds earlier in this scenario than in post-inflationary structure formation without a resonant phase~\cite{Musoke2019,Eggemeier2020,Eggemeier2021,Eggemeier2022}, significantly extending the possible era of post-inflationary structure formation, as well as changing the form of the inflaton HMF. 
The resonant phase itself can generate a high frequency gravitational wave background~\cite{Khlebnikov1997_grav,Easther2006_1,Easther2006_2,Easther2007,Garcia-Bellido2007_2,Dufaux2007,Dufaux2008,Zhou2013, Amin2018,Hiramatsu2020,Lozanov2022} but this will be diluted during a long post-inflationary matter-dominated phase.  However, nonlinear dynamics during the matter-dominated phase will also source a gravitational wave background, but at lower frequencies~\cite{Eggemeier2022}. Given that resonance can significantly extend the duration of the nonlinear phase it is plausible that it will also boost the resulting gravitational wave signal. 
Similarly, any PBH production can be enhanced by the longer nonlinear phase~\cite{Eggemeier2021,Padilla2022,Hidalgo2022}. 
 
We have focused on a  model with comparatively mild parametric resonance, forestalling the formation of oscillons. Once formed, oscillons decouple from the cosmological expansion, so the spatial resolution would need to be increased continuously in the Klein-Gordon simulation to capture their dynamics and  oscillons are frequently long-lived relative to the Hubble time. Some oscillons can be captured in a Schrödinger-Poisson framework if self-interactions are included~\cite{Amin:2019ums} but this is not generically possible. As in simulations of the early axion evolution~\cite{Vaquero2018,Buschmann2019,OHare2021,Eggemeier2022_voids}, the oscillons would then need to evaporate and the inflaton field would become nonrelativistic before the current gravitational solver can be applied.  

Relative to gravitational time scales $\delta \rho / \rho $ grows very rapidly during resonance. For the scenario here it peaks at $\mathcal{O}(10^{-1})$ during this phase, so we can be confident that gravitational effects during resonance are small and the division between the Klein-Gordon and Schrödinger-Poisson regimes is clear-cut. 
Nevertheless, our  results might be weakly sensitive to the point of transition from the Klein-Gordon to the Schrödinger-Poisson phase. 

There are clearly a number of avenues for follow-up, the first and most obvious is to compute the HMF at larger values of $\mathcal{N}$. This might be achieved by Press-Schechter-style semianalytic calculations~\cite{Niemeyer2019} or using the mapping  between the wave function and an $N$-body representation developed for the halo finding calculation to initialize an  $N$-body solver. This  would exploit the known correspondence between  $N$-body and wave function descriptions of wave-like matter (whether inflaton or FDM) at scales where interference effects are not relevant~\cite{Eggemeier2020}. With this information in hand we could then accurately estimate the SGWB produced by this scenario. 

So far as  we are aware, these simulations appear to yield the least massive gravitationally bound objects that have been discussed numerically within a weak-field cosmological framework. Moreover, this work further develops the parallel treatment of nonlinear gravitational dynamics in the post-inflationary universe and ``present epoch'' structure formation first described in Refs.~\cite{Musoke2019,Niemeyer2019}.

That said, we clearly want to be able to extend this treatment to a broad range of inflationary scenarios, including those with couplings to other fields or which support oscillon production. Likewise, it is conceivable that gravitational interactions during a lengthy oscillon-dominated phase will be significant. Some investigations of gravitational effects on reheating and oscillon dynamics have been made~\cite{Giblin:2019nuv,Kou:2019bbc,Kou:2021bij,Aurrekoetxea:2023jwd} but these are omitted from most cosmological Klein-Gordon solvers, and this points to a new range of possible dynamics. Conversely, Amin and Mocz~\cite{Amin:2019ums} investigated the gravitational dynamics of oscillons in an expanding universe by adding interaction terms to the Schrödinger equation but this approach need not capture the full behaviour that would be revealed by a Klein-Gordon solver.
Finally, this analysis and almost all similar treatments assumes that thermalization eventually takes place but does not provide a detailed description of the process or the ways in which the detailed couplings involved modify the latter stages of the matter-dominated phase. Consequently, there are many open and interesting questions to pursue and answering them will allow us to build a complete understanding of the primordial universe. 

\vspace*{0.5cm}

\section*{Acknowledgements}
We thank Mateja Gosenca, Emily Kendall, Yourong (Frank) Wang, and Angela Xue for useful discussions. 
BE gratefully acknowledges the computing time provided on the supercomputers Lise and Emmy at NHR@ZIB and NHR@Göttingen as part of the NHR infrastructure. We acknowledge the yt toolkit~\cite{yt} that was used for the analysis of numerical data. PH and RE acknowledge support from the Marsden Fund of the Royal Society of New Zealand. This collaboration was supported by a Julius von Haast Fellowship Award provided by the New Zealand Ministry of Business, Innovation and Employment  administered by the Royal Society of New Zealand.

\bibliography{refs}

\begin{thebibliography}{90}%
\makeatletter
\providecommand \@ifxundefined [1]{%
 \@ifx{#1\undefined}
}%
\providecommand \@ifnum [1]{%
 \ifnum #1\expandafter \@firstoftwo
 \else \expandafter \@secondoftwo
 \fi
}%
\providecommand \@ifx [1]{%
 \ifx #1\expandafter \@firstoftwo
 \else \expandafter \@secondoftwo
 \fi
}%
\providecommand \natexlab [1]{#1}%
\providecommand \enquote  [1]{``#1''}%
\providecommand \bibnamefont  [1]{#1}%
\providecommand \bibfnamefont [1]{#1}%
\providecommand \citenamefont [1]{#1}%
\providecommand \href@noop [0]{\@secondoftwo}%
\providecommand \href [0]{\begingroup \@sanitize@url \@href}%
\providecommand \@href[1]{\@@startlink{#1}\@@href}%
\providecommand \@@href[1]{\endgroup#1\@@endlink}%
\providecommand \@sanitize@url [0]{\catcode `\\12\catcode `\$12\catcode
  `\&12\catcode `\#12\catcode `\^12\catcode `\_12\catcode `\%12\relax}%
\providecommand \@@startlink[1]{}%
\providecommand \@@endlink[0]{}%
\providecommand \url  [0]{\begingroup\@sanitize@url \@url }%
\providecommand \@url [1]{\endgroup\@href {#1}{\urlprefix }}%
\providecommand \urlprefix  [0]{URL }%
\providecommand \Eprint [0]{\href }%
\providecommand \doibase [0]{http://dx.doi.org/}%
\providecommand \selectlanguage [0]{\@gobble}%
\providecommand \bibinfo  [0]{\@secondoftwo}%
\providecommand \bibfield  [0]{\@secondoftwo}%
\providecommand \translation [1]{[#1]}%
\providecommand \BibitemOpen [0]{}%
\providecommand \bibitemStop [0]{}%
\providecommand \bibitemNoStop [0]{.\EOS\space}%
\providecommand \EOS [0]{\spacefactor3000\relax}%
\providecommand \BibitemShut  [1]{\csname bibitem#1\endcsname}%
\let\auto@bib@innerbib\@empty
\bibitem [{\citenamefont {Starobinsky}(1980)}]{Starobinsky1980}%
  \BibitemOpen
  \bibfield  {author} {\bibinfo {author} {\bibfnamefont {A.}~\bibnamefont
  {Starobinsky}},\ }\href {\doibase
  https://doi.org/10.1016/0370-2693(80)90670-X} {\bibfield  {journal} {\bibinfo
   {journal} {Physics Letters B}\ }\textbf {\bibinfo {volume} {91}},\ \bibinfo
  {pages} {99 } (\bibinfo {year} {1980})}\BibitemShut {NoStop}%
\bibitem [{\citenamefont {Guth}(1981)}]{Guth1981}%
  \BibitemOpen
  \bibfield  {author} {\bibinfo {author} {\bibfnamefont {A.~H.}\ \bibnamefont
  {Guth}},\ }\href {\doibase 10.1103/PhysRevD.23.347} {\bibfield  {journal}
  {\bibinfo  {journal} {Phys. Rev. D}\ }\textbf {\bibinfo {volume} {23}},\
  \bibinfo {pages} {347} (\bibinfo {year} {1981})}\BibitemShut {NoStop}%
\bibitem [{\citenamefont {Linde}(1982)}]{Linde1982}%
  \BibitemOpen
  \bibfield  {author} {\bibinfo {author} {\bibfnamefont {A.}~\bibnamefont
  {Linde}},\ }\href {\doibase https://doi.org/10.1016/0370-2693(82)91219-9}
  {\bibfield  {journal} {\bibinfo  {journal} {Physics Letters B}\ }\textbf
  {\bibinfo {volume} {108}},\ \bibinfo {pages} {389 } (\bibinfo {year}
  {1982})}\BibitemShut {NoStop}%
\bibitem [{\citenamefont {Linde}(1983)}]{Linde1983}%
  \BibitemOpen
  \bibfield  {author} {\bibinfo {author} {\bibfnamefont {A.}~\bibnamefont
  {Linde}},\ }\href {\doibase https://doi.org/10.1016/0370-2693(83)90837-7}
  {\bibfield  {journal} {\bibinfo  {journal} {Physics Letters B}\ }\textbf
  {\bibinfo {volume} {129}},\ \bibinfo {pages} {177 } (\bibinfo {year}
  {1983})}\BibitemShut {NoStop}%
\bibitem [{\citenamefont {Kawasaki}\ \emph {et~al.}(1999)\citenamefont
  {Kawasaki}, \citenamefont {Kohri},\ and\ \citenamefont
  {Sugiyama}}]{Kawasaki1999}%
  \BibitemOpen
  \bibfield  {author} {\bibinfo {author} {\bibfnamefont {M.}~\bibnamefont
  {Kawasaki}}, \bibinfo {author} {\bibfnamefont {K.}~\bibnamefont {Kohri}}, \
  and\ \bibinfo {author} {\bibfnamefont {N.}~\bibnamefont {Sugiyama}},\ }\href
  {\doibase 10.1103/PhysRevLett.82.4168} {\bibfield  {journal} {\bibinfo
  {journal} {Phys. Rev. Lett.}\ }\textbf {\bibinfo {volume} {82}},\ \bibinfo
  {pages} {4168} (\bibinfo {year} {1999})}\BibitemShut {NoStop}%
\bibitem [{\citenamefont {Hannestad}(2004)}]{Hannestad2004}%
  \BibitemOpen
  \bibfield  {author} {\bibinfo {author} {\bibfnamefont {S.}~\bibnamefont
  {Hannestad}},\ }\href {\doibase 10.1103/PhysRevD.70.043506} {\bibfield
  {journal} {\bibinfo  {journal} {Phys. Rev. D}\ }\textbf {\bibinfo {volume}
  {70}},\ \bibinfo {pages} {043506} (\bibinfo {year} {2004})}\BibitemShut
  {NoStop}%
\bibitem [{\citenamefont {de~Salas}\ \emph {et~al.}(2015)\citenamefont
  {de~Salas}, \citenamefont {Lattanzi}, \citenamefont {Mangano}, \citenamefont
  {Miele}, \citenamefont {Pastor},\ and\ \citenamefont {Pisanti}}]{Salas2015}%
  \BibitemOpen
  \bibfield  {author} {\bibinfo {author} {\bibfnamefont {P.~F.}\ \bibnamefont
  {de~Salas}}, \bibinfo {author} {\bibfnamefont {M.}~\bibnamefont {Lattanzi}},
  \bibinfo {author} {\bibfnamefont {G.}~\bibnamefont {Mangano}}, \bibinfo
  {author} {\bibfnamefont {G.}~\bibnamefont {Miele}}, \bibinfo {author}
  {\bibfnamefont {S.}~\bibnamefont {Pastor}}, \ and\ \bibinfo {author}
  {\bibfnamefont {O.}~\bibnamefont {Pisanti}},\ }\href {\doibase
  10.1103/PhysRevD.92.123534} {\bibfield  {journal} {\bibinfo  {journal} {Phys.
  Rev. D}\ }\textbf {\bibinfo {volume} {92}},\ \bibinfo {pages} {123534}
  (\bibinfo {year} {2015})}\BibitemShut {NoStop}%
\bibitem [{\citenamefont {Traschen}\ and\ \citenamefont
  {Brandenberger}(1990)}]{Traschen1990}%
  \BibitemOpen
  \bibfield  {author} {\bibinfo {author} {\bibfnamefont {J.~H.}\ \bibnamefont
  {Traschen}}\ and\ \bibinfo {author} {\bibfnamefont {R.~H.}\ \bibnamefont
  {Brandenberger}},\ }\href {\doibase 10.1103/PhysRevD.42.2491} {\bibfield
  {journal} {\bibinfo  {journal} {Phys. Rev. D}\ }\textbf {\bibinfo {volume}
  {42}},\ \bibinfo {pages} {2491} (\bibinfo {year} {1990})}\BibitemShut
  {NoStop}%
\bibitem [{\citenamefont {Kofman}\ \emph {et~al.}(1994)\citenamefont {Kofman},
  \citenamefont {Linde},\ and\ \citenamefont {Starobinsky}}]{Kofman1994}%
  \BibitemOpen
  \bibfield  {author} {\bibinfo {author} {\bibfnamefont {L.}~\bibnamefont
  {Kofman}}, \bibinfo {author} {\bibfnamefont {A.}~\bibnamefont {Linde}}, \
  and\ \bibinfo {author} {\bibfnamefont {A.~A.}\ \bibnamefont {Starobinsky}},\
  }\href {\doibase 10.1103/PhysRevLett.73.3195} {\bibfield  {journal} {\bibinfo
   {journal} {Phys. Rev. Lett.}\ }\textbf {\bibinfo {volume} {73}},\ \bibinfo
  {pages} {3195} (\bibinfo {year} {1994})}\BibitemShut {NoStop}%
\bibitem [{\citenamefont {Shtanov}\ \emph {et~al.}(1995)\citenamefont
  {Shtanov}, \citenamefont {Traschen},\ and\ \citenamefont
  {Brandenberger}}]{Shtanov1994}%
  \BibitemOpen
  \bibfield  {author} {\bibinfo {author} {\bibfnamefont {Y.}~\bibnamefont
  {Shtanov}}, \bibinfo {author} {\bibfnamefont {J.~H.}\ \bibnamefont
  {Traschen}}, \ and\ \bibinfo {author} {\bibfnamefont {R.~H.}\ \bibnamefont
  {Brandenberger}},\ }\href {\doibase 10.1103/PhysRevD.51.5438} {\bibfield
  {journal} {\bibinfo  {journal} {Phys. Rev.}\ }\textbf {\bibinfo {volume}
  {D51}},\ \bibinfo {pages} {5438} (\bibinfo {year} {1995})},\ \Eprint
  {http://arxiv.org/abs/hep-ph/9407247} {arXiv:hep-ph/9407247 [hep-ph]}
  \BibitemShut {NoStop}%
\bibitem [{\citenamefont {Kofman}\ \emph {et~al.}(1997)\citenamefont {Kofman},
  \citenamefont {Linde},\ and\ \citenamefont {Starobinsky}}]{Kofman1997}%
  \BibitemOpen
  \bibfield  {author} {\bibinfo {author} {\bibfnamefont {L.}~\bibnamefont
  {Kofman}}, \bibinfo {author} {\bibfnamefont {A.}~\bibnamefont {Linde}}, \
  and\ \bibinfo {author} {\bibfnamefont {A.~A.}\ \bibnamefont {Starobinsky}},\
  }\href {\doibase 10.1103/PhysRevD.56.3258} {\bibfield  {journal} {\bibinfo
  {journal} {Phys. Rev. D}\ }\textbf {\bibinfo {volume} {56}},\ \bibinfo
  {pages} {3258} (\bibinfo {year} {1997})}\BibitemShut {NoStop}%
\bibitem [{\citenamefont {Akrami}\ \emph {et~al.}(2020)\citenamefont {Akrami}
  \emph {et~al.}}]{Planck2018_inflation}%
  \BibitemOpen
  \bibfield  {author} {\bibinfo {author} {\bibfnamefont {Y.}~\bibnamefont
  {Akrami}} \emph {et~al.} (\bibinfo {collaboration} {Planck}),\ }\href
  {\doibase 10.1051/0004-6361/201833887} {\bibfield  {journal} {\bibinfo
  {journal} {Astron. Astrophys.}\ }\textbf {\bibinfo {volume} {641}},\ \bibinfo
  {pages} {A10} (\bibinfo {year} {2020})},\ \Eprint
  {http://arxiv.org/abs/1807.06211} {arXiv:1807.06211 [astro-ph.CO]}
  \BibitemShut {NoStop}%
\bibitem [{\citenamefont {Ade}\ \emph {et~al.}(2021)\citenamefont {Ade} \emph
  {et~al.}}]{BICEP:2021xfz}%
  \BibitemOpen
  \bibfield  {author} {\bibinfo {author} {\bibfnamefont {P.~A.~R.}\
  \bibnamefont {Ade}} \emph {et~al.} (\bibinfo {collaboration} {BICEP, Keck}),\
  }\href {\doibase 10.1103/PhysRevLett.127.151301} {\bibfield  {journal}
  {\bibinfo  {journal} {Phys. Rev. Lett.}\ }\textbf {\bibinfo {volume} {127}},\
  \bibinfo {pages} {151301} (\bibinfo {year} {2021})},\ \Eprint
  {http://arxiv.org/abs/2110.00483} {arXiv:2110.00483 [astro-ph.CO]}
  \BibitemShut {NoStop}%
\bibitem [{\citenamefont {Gleiser}(1994)}]{Gleiser1993}%
  \BibitemOpen
  \bibfield  {author} {\bibinfo {author} {\bibfnamefont {M.}~\bibnamefont
  {Gleiser}},\ }\href {\doibase 10.1103/PhysRevD.49.2978} {\bibfield  {journal}
  {\bibinfo  {journal} {Phys. Rev. D}\ }\textbf {\bibinfo {volume} {49}},\
  \bibinfo {pages} {2978} (\bibinfo {year} {1994})},\ \Eprint
  {http://arxiv.org/abs/hep-ph/9308279} {arXiv:hep-ph/9308279} \BibitemShut
  {NoStop}%
\bibitem [{\citenamefont {Copeland}\ \emph {et~al.}(1995)\citenamefont
  {Copeland}, \citenamefont {Gleiser},\ and\ \citenamefont
  {Muller}}]{Copeland1995}%
  \BibitemOpen
  \bibfield  {author} {\bibinfo {author} {\bibfnamefont {E.~J.}\ \bibnamefont
  {Copeland}}, \bibinfo {author} {\bibfnamefont {M.}~\bibnamefont {Gleiser}}, \
  and\ \bibinfo {author} {\bibfnamefont {H.~R.}\ \bibnamefont {Muller}},\
  }\href {\doibase 10.1103/PhysRevD.52.1920} {\bibfield  {journal} {\bibinfo
  {journal} {Phys. Rev. D}\ }\textbf {\bibinfo {volume} {52}},\ \bibinfo
  {pages} {1920} (\bibinfo {year} {1995})},\ \Eprint
  {http://arxiv.org/abs/hep-ph/9503217} {arXiv:hep-ph/9503217} \BibitemShut
  {NoStop}%
\bibitem [{\citenamefont {Amin}\ \emph {et~al.}(2010)\citenamefont {Amin},
  \citenamefont {Easther},\ and\ \citenamefont {Finkel}}]{Amin2010}%
  \BibitemOpen
  \bibfield  {author} {\bibinfo {author} {\bibfnamefont {M.~A.}\ \bibnamefont
  {Amin}}, \bibinfo {author} {\bibfnamefont {R.}~\bibnamefont {Easther}}, \
  and\ \bibinfo {author} {\bibfnamefont {H.}~\bibnamefont {Finkel}},\ }\href
  {\doibase 10.1088/1475-7516/2010/12/001} {\bibfield  {journal} {\bibinfo
  {journal} {JCAP}\ }\textbf {\bibinfo {volume} {12}},\ \bibinfo {pages} {001}
  (\bibinfo {year} {2010})},\ \Eprint {http://arxiv.org/abs/1009.2505}
  {arXiv:1009.2505 [astro-ph.CO]} \BibitemShut {NoStop}%
\bibitem [{\citenamefont {Amin}\ \emph {et~al.}(2012)\citenamefont {Amin},
  \citenamefont {Easther}, \citenamefont {Finkel}, \citenamefont {Flauger},\
  and\ \citenamefont {Hertzberg}}]{Amin2012}%
  \BibitemOpen
  \bibfield  {author} {\bibinfo {author} {\bibfnamefont {M.~A.}\ \bibnamefont
  {Amin}}, \bibinfo {author} {\bibfnamefont {R.}~\bibnamefont {Easther}},
  \bibinfo {author} {\bibfnamefont {H.}~\bibnamefont {Finkel}}, \bibinfo
  {author} {\bibfnamefont {R.}~\bibnamefont {Flauger}}, \ and\ \bibinfo
  {author} {\bibfnamefont {M.~P.}\ \bibnamefont {Hertzberg}},\ }\href {\doibase
  10.1103/PhysRevLett.108.241302} {\bibfield  {journal} {\bibinfo  {journal}
  {Phys. Rev. Lett.}\ }\textbf {\bibinfo {volume} {108}},\ \bibinfo {pages}
  {241302} (\bibinfo {year} {2012})}\BibitemShut {NoStop}%
\bibitem [{\citenamefont {Lozanov}\ and\ \citenamefont
  {Amin}(2018)}]{Lozanov2017}%
  \BibitemOpen
  \bibfield  {author} {\bibinfo {author} {\bibfnamefont {K.~D.}\ \bibnamefont
  {Lozanov}}\ and\ \bibinfo {author} {\bibfnamefont {M.~A.}\ \bibnamefont
  {Amin}},\ }\href {\doibase 10.1103/PhysRevD.97.023533} {\bibfield  {journal}
  {\bibinfo  {journal} {Phys. Rev. D}\ }\textbf {\bibinfo {volume} {97}},\
  \bibinfo {pages} {023533} (\bibinfo {year} {2018})},\ \Eprint
  {http://arxiv.org/abs/1710.06851} {arXiv:1710.06851 [astro-ph.CO]}
  \BibitemShut {NoStop}%
\bibitem [{\citenamefont {Khlebnikov}\ and\ \citenamefont
  {Tkachev}(1997)}]{Khlebnikov1997_grav}%
  \BibitemOpen
  \bibfield  {author} {\bibinfo {author} {\bibfnamefont {S.~Y.}\ \bibnamefont
  {Khlebnikov}}\ and\ \bibinfo {author} {\bibfnamefont {I.~I.}\ \bibnamefont
  {Tkachev}},\ }\href {\doibase 10.1103/PhysRevD.56.653} {\bibfield  {journal}
  {\bibinfo  {journal} {Phys. Rev. D}\ }\textbf {\bibinfo {volume} {56}},\
  \bibinfo {pages} {653} (\bibinfo {year} {1997})},\ \Eprint
  {http://arxiv.org/abs/hep-ph/9701423} {arXiv:hep-ph/9701423} \BibitemShut
  {NoStop}%
\bibitem [{\citenamefont {Easther}\ and\ \citenamefont
  {Lim}(2006)}]{Easther2006_1}%
  \BibitemOpen
  \bibfield  {author} {\bibinfo {author} {\bibfnamefont {R.}~\bibnamefont
  {Easther}}\ and\ \bibinfo {author} {\bibfnamefont {E.~A.}\ \bibnamefont
  {Lim}},\ }\href {\doibase 10.1088/1475-7516/2006/04/010} {\bibfield
  {journal} {\bibinfo  {journal} {JCAP}\ }\textbf {\bibinfo {volume} {04}},\
  \bibinfo {pages} {010} (\bibinfo {year} {2006})},\ \Eprint
  {http://arxiv.org/abs/astro-ph/0601617} {arXiv:astro-ph/0601617} \BibitemShut
  {NoStop}%
\bibitem [{\citenamefont {Easther}\ \emph {et~al.}(2007)\citenamefont
  {Easther}, \citenamefont {Giblin},\ and\ \citenamefont
  {Lim}}]{Easther2006_2}%
  \BibitemOpen
  \bibfield  {author} {\bibinfo {author} {\bibfnamefont {R.}~\bibnamefont
  {Easther}}, \bibinfo {author} {\bibfnamefont {J.~T.}\ \bibnamefont {Giblin},
  \bibfnamefont {Jr.}}, \ and\ \bibinfo {author} {\bibfnamefont {E.~A.}\
  \bibnamefont {Lim}},\ }\href {\doibase 10.1103/PhysRevLett.99.221301}
  {\bibfield  {journal} {\bibinfo  {journal} {Phys. Rev. Lett.}\ }\textbf
  {\bibinfo {volume} {99}},\ \bibinfo {pages} {221301} (\bibinfo {year}
  {2007})},\ \Eprint {http://arxiv.org/abs/astro-ph/0612294}
  {arXiv:astro-ph/0612294} \BibitemShut {NoStop}%
\bibitem [{\citenamefont {Easther}\ \emph {et~al.}(2008)\citenamefont
  {Easther}, \citenamefont {Giblin},\ and\ \citenamefont {Lim}}]{Easther2007}%
  \BibitemOpen
  \bibfield  {author} {\bibinfo {author} {\bibfnamefont {R.}~\bibnamefont
  {Easther}}, \bibinfo {author} {\bibfnamefont {J.~T.}\ \bibnamefont {Giblin}},
  \ and\ \bibinfo {author} {\bibfnamefont {E.~A.}\ \bibnamefont {Lim}},\ }\href
  {\doibase 10.1103/PhysRevD.77.103519} {\bibfield  {journal} {\bibinfo
  {journal} {Phys. Rev. D}\ }\textbf {\bibinfo {volume} {77}},\ \bibinfo
  {pages} {103519} (\bibinfo {year} {2008})},\ \Eprint
  {http://arxiv.org/abs/0712.2991} {arXiv:0712.2991 [astro-ph]} \BibitemShut
  {NoStop}%
\bibitem [{\citenamefont {Garcia-Bellido}\ \emph {et~al.}(2008)\citenamefont
  {Garcia-Bellido}, \citenamefont {Figueroa},\ and\ \citenamefont
  {Sastre}}]{Garcia-Bellido2007_2}%
  \BibitemOpen
  \bibfield  {author} {\bibinfo {author} {\bibfnamefont {J.}~\bibnamefont
  {Garcia-Bellido}}, \bibinfo {author} {\bibfnamefont {D.~G.}\ \bibnamefont
  {Figueroa}}, \ and\ \bibinfo {author} {\bibfnamefont {A.}~\bibnamefont
  {Sastre}},\ }\href {\doibase 10.1103/PhysRevD.77.043517} {\bibfield
  {journal} {\bibinfo  {journal} {Phys. Rev. D}\ }\textbf {\bibinfo {volume}
  {77}},\ \bibinfo {pages} {043517} (\bibinfo {year} {2008})},\ \Eprint
  {http://arxiv.org/abs/0707.0839} {arXiv:0707.0839 [hep-ph]} \BibitemShut
  {NoStop}%
\bibitem [{\citenamefont {Dufaux}\ \emph {et~al.}(2007)\citenamefont {Dufaux},
  \citenamefont {Bergman}, \citenamefont {Felder}, \citenamefont {Kofman},\
  and\ \citenamefont {Uzan}}]{Dufaux2007}%
  \BibitemOpen
  \bibfield  {author} {\bibinfo {author} {\bibfnamefont {J.~F.}\ \bibnamefont
  {Dufaux}}, \bibinfo {author} {\bibfnamefont {A.}~\bibnamefont {Bergman}},
  \bibinfo {author} {\bibfnamefont {G.~N.}\ \bibnamefont {Felder}}, \bibinfo
  {author} {\bibfnamefont {L.}~\bibnamefont {Kofman}}, \ and\ \bibinfo {author}
  {\bibfnamefont {J.-P.}\ \bibnamefont {Uzan}},\ }\href {\doibase
  10.1103/PhysRevD.76.123517} {\bibfield  {journal} {\bibinfo  {journal} {Phys.
  Rev. D}\ }\textbf {\bibinfo {volume} {76}},\ \bibinfo {pages} {123517}
  (\bibinfo {year} {2007})},\ \Eprint {http://arxiv.org/abs/0707.0875}
  {arXiv:0707.0875 [astro-ph]} \BibitemShut {NoStop}%
\bibitem [{\citenamefont {Dufaux}\ \emph {et~al.}(2009)\citenamefont {Dufaux},
  \citenamefont {Felder}, \citenamefont {Kofman},\ and\ \citenamefont
  {Navros}}]{Dufaux2008}%
  \BibitemOpen
  \bibfield  {author} {\bibinfo {author} {\bibfnamefont {J.-F.}\ \bibnamefont
  {Dufaux}}, \bibinfo {author} {\bibfnamefont {G.}~\bibnamefont {Felder}},
  \bibinfo {author} {\bibfnamefont {L.}~\bibnamefont {Kofman}}, \ and\ \bibinfo
  {author} {\bibfnamefont {O.}~\bibnamefont {Navros}},\ }\href {\doibase
  10.1088/1475-7516/2009/03/001} {\bibfield  {journal} {\bibinfo  {journal}
  {JCAP}\ }\textbf {\bibinfo {volume} {03}},\ \bibinfo {pages} {001} (\bibinfo
  {year} {2009})},\ \Eprint {http://arxiv.org/abs/0812.2917} {arXiv:0812.2917
  [astro-ph]} \BibitemShut {NoStop}%
\bibitem [{\citenamefont {Zhou}\ \emph {et~al.}(2013)\citenamefont {Zhou},
  \citenamefont {Copeland}, \citenamefont {Easther}, \citenamefont {Finkel},
  \citenamefont {Mou},\ and\ \citenamefont {Saffin}}]{Zhou2013}%
  \BibitemOpen
  \bibfield  {author} {\bibinfo {author} {\bibfnamefont {S.-Y.}\ \bibnamefont
  {Zhou}}, \bibinfo {author} {\bibfnamefont {E.~J.}\ \bibnamefont {Copeland}},
  \bibinfo {author} {\bibfnamefont {R.}~\bibnamefont {Easther}}, \bibinfo
  {author} {\bibfnamefont {H.}~\bibnamefont {Finkel}}, \bibinfo {author}
  {\bibfnamefont {Z.-G.}\ \bibnamefont {Mou}}, \ and\ \bibinfo {author}
  {\bibfnamefont {P.~M.}\ \bibnamefont {Saffin}},\ }\href {\doibase
  10.1007/JHEP10(2013)026} {\bibfield  {journal} {\bibinfo  {journal} {JHEP}\
  }\textbf {\bibinfo {volume} {10}},\ \bibinfo {pages} {026} (\bibinfo {year}
  {2013})},\ \Eprint {http://arxiv.org/abs/1304.6094} {arXiv:1304.6094
  [astro-ph.CO]} \BibitemShut {NoStop}%
\bibitem [{\citenamefont {Amin}\ \emph {et~al.}(2018)\citenamefont {Amin},
  \citenamefont {Braden}, \citenamefont {Copeland}, \citenamefont {Giblin},
  \citenamefont {Solorio}, \citenamefont {Weiner},\ and\ \citenamefont
  {Zhou}}]{Amin2018}%
  \BibitemOpen
  \bibfield  {author} {\bibinfo {author} {\bibfnamefont {M.~A.}\ \bibnamefont
  {Amin}}, \bibinfo {author} {\bibfnamefont {J.}~\bibnamefont {Braden}},
  \bibinfo {author} {\bibfnamefont {E.~J.}\ \bibnamefont {Copeland}}, \bibinfo
  {author} {\bibfnamefont {J.~T.}\ \bibnamefont {Giblin}}, \bibinfo {author}
  {\bibfnamefont {C.}~\bibnamefont {Solorio}}, \bibinfo {author} {\bibfnamefont
  {Z.~J.}\ \bibnamefont {Weiner}}, \ and\ \bibinfo {author} {\bibfnamefont
  {S.-Y.}\ \bibnamefont {Zhou}},\ }\href {\doibase 10.1103/PhysRevD.98.024040}
  {\bibfield  {journal} {\bibinfo  {journal} {Phys. Rev. D}\ }\textbf {\bibinfo
  {volume} {98}},\ \bibinfo {pages} {024040} (\bibinfo {year} {2018})},\
  \Eprint {http://arxiv.org/abs/1803.08047} {arXiv:1803.08047 [astro-ph.CO]}
  \BibitemShut {NoStop}%
\bibitem [{\citenamefont {Hiramatsu}\ \emph {et~al.}(2021)\citenamefont
  {Hiramatsu}, \citenamefont {Sfakianakis},\ and\ \citenamefont
  {Yamaguchi}}]{Hiramatsu2020}%
  \BibitemOpen
  \bibfield  {author} {\bibinfo {author} {\bibfnamefont {T.}~\bibnamefont
  {Hiramatsu}}, \bibinfo {author} {\bibfnamefont {E.~I.}\ \bibnamefont
  {Sfakianakis}}, \ and\ \bibinfo {author} {\bibfnamefont {M.}~\bibnamefont
  {Yamaguchi}},\ }\href {\doibase 10.1007/JHEP03(2021)021} {\bibfield
  {journal} {\bibinfo  {journal} {JHEP}\ }\textbf {\bibinfo {volume} {03}},\
  \bibinfo {pages} {021} (\bibinfo {year} {2021})},\ \Eprint
  {http://arxiv.org/abs/2011.12201} {arXiv:2011.12201 [hep-ph]} \BibitemShut
  {NoStop}%
\bibitem [{\citenamefont {Lozanov}\ and\ \citenamefont
  {Takhistov}(2023)}]{Lozanov2022}%
  \BibitemOpen
  \bibfield  {author} {\bibinfo {author} {\bibfnamefont {K.~D.}\ \bibnamefont
  {Lozanov}}\ and\ \bibinfo {author} {\bibfnamefont {V.}~\bibnamefont
  {Takhistov}},\ }\href {\doibase 10.1103/PhysRevLett.130.181002} {\bibfield
  {journal} {\bibinfo  {journal} {Phys. Rev. Lett.}\ }\textbf {\bibinfo
  {volume} {130}},\ \bibinfo {pages} {181002} (\bibinfo {year} {2023})},\
  \Eprint {http://arxiv.org/abs/2204.07152} {arXiv:2204.07152 [astro-ph.CO]}
  \BibitemShut {NoStop}%
\bibitem [{\citenamefont {Cotner}\ \emph {et~al.}(2018)\citenamefont {Cotner},
  \citenamefont {Kusenko},\ and\ \citenamefont {Takhistov}}]{Cotner2018}%
  \BibitemOpen
  \bibfield  {author} {\bibinfo {author} {\bibfnamefont {E.}~\bibnamefont
  {Cotner}}, \bibinfo {author} {\bibfnamefont {A.}~\bibnamefont {Kusenko}}, \
  and\ \bibinfo {author} {\bibfnamefont {V.}~\bibnamefont {Takhistov}},\ }\href
  {\doibase 10.1103/PhysRevD.98.083513} {\bibfield  {journal} {\bibinfo
  {journal} {Phys. Rev. D}\ }\textbf {\bibinfo {volume} {98}},\ \bibinfo
  {pages} {083513} (\bibinfo {year} {2018})},\ \Eprint
  {http://arxiv.org/abs/1801.03321} {arXiv:1801.03321 [astro-ph.CO]}
  \BibitemShut {NoStop}%
\bibitem [{\citenamefont {Jedamzik}\ \emph
  {et~al.}(2010{\natexlab{a}})\citenamefont {Jedamzik}, \citenamefont
  {Lemoine},\ and\ \citenamefont {Martin}}]{Jedamzik2010_dens}%
  \BibitemOpen
  \bibfield  {author} {\bibinfo {author} {\bibfnamefont {K.}~\bibnamefont
  {Jedamzik}}, \bibinfo {author} {\bibfnamefont {M.}~\bibnamefont {Lemoine}}, \
  and\ \bibinfo {author} {\bibfnamefont {J.}~\bibnamefont {Martin}},\ }\href
  {\doibase 10.1088/1475-7516/2010/09/034} {\bibfield  {journal} {\bibinfo
  {journal} {JCAP}\ }\textbf {\bibinfo {volume} {09}},\ \bibinfo {pages} {034}
  (\bibinfo {year} {2010}{\natexlab{a}})},\ \Eprint
  {http://arxiv.org/abs/1002.3039} {arXiv:1002.3039 [astro-ph.CO]} \BibitemShut
  {NoStop}%
\bibitem [{\citenamefont {Easther}\ \emph {et~al.}(2011)\citenamefont
  {Easther}, \citenamefont {Flauger},\ and\ \citenamefont
  {Gilmore}}]{Easther2010}%
  \BibitemOpen
  \bibfield  {author} {\bibinfo {author} {\bibfnamefont {R.}~\bibnamefont
  {Easther}}, \bibinfo {author} {\bibfnamefont {R.}~\bibnamefont {Flauger}}, \
  and\ \bibinfo {author} {\bibfnamefont {J.~B.}\ \bibnamefont {Gilmore}},\
  }\href {\doibase 10.1088/1475-7516/2011/04/027} {\bibfield  {journal}
  {\bibinfo  {journal} {JCAP}\ }\textbf {\bibinfo {volume} {1104}},\ \bibinfo
  {pages} {027} (\bibinfo {year} {2011})},\ \Eprint
  {http://arxiv.org/abs/1003.3011} {arXiv:1003.3011 [astro-ph.CO]} \BibitemShut
  {NoStop}%
\bibitem [{\citenamefont {Musoke}\ \emph {et~al.}(2020)\citenamefont {Musoke},
  \citenamefont {Hotchkiss},\ and\ \citenamefont {Easther}}]{Musoke2019}%
  \BibitemOpen
  \bibfield  {author} {\bibinfo {author} {\bibfnamefont {N.}~\bibnamefont
  {Musoke}}, \bibinfo {author} {\bibfnamefont {S.}~\bibnamefont {Hotchkiss}}, \
  and\ \bibinfo {author} {\bibfnamefont {R.}~\bibnamefont {Easther}},\ }\href
  {\doibase 10.1103/PhysRevLett.124.061301} {\bibfield  {journal} {\bibinfo
  {journal} {Phys. Rev. Lett.}\ }\textbf {\bibinfo {volume} {124}},\ \bibinfo
  {pages} {061301} (\bibinfo {year} {2020})},\ \Eprint
  {http://arxiv.org/abs/1909.11678} {arXiv:1909.11678 [astro-ph.CO]}
  \BibitemShut {NoStop}%
\bibitem [{\citenamefont {Niemeyer}\ and\ \citenamefont
  {Easther}(2020)}]{Niemeyer2019}%
  \BibitemOpen
  \bibfield  {author} {\bibinfo {author} {\bibfnamefont {J.~C.}\ \bibnamefont
  {Niemeyer}}\ and\ \bibinfo {author} {\bibfnamefont {R.}~\bibnamefont
  {Easther}},\ }\href {\doibase 10.1088/1475-7516/2020/07/030} {\bibfield
  {journal} {\bibinfo  {journal} {Journal of Cosmology and Astroparticle
  Physics}\ }\textbf {\bibinfo {volume} {2020}},\ \bibinfo {pages} {030}
  (\bibinfo {year} {2020})}\BibitemShut {NoStop}%
\bibitem [{\citenamefont {Eggemeier}\ \emph {et~al.}(2021)\citenamefont
  {Eggemeier}, \citenamefont {Niemeyer},\ and\ \citenamefont
  {Easther}}]{Eggemeier2020}%
  \BibitemOpen
  \bibfield  {author} {\bibinfo {author} {\bibfnamefont {B.}~\bibnamefont
  {Eggemeier}}, \bibinfo {author} {\bibfnamefont {J.~C.}\ \bibnamefont
  {Niemeyer}}, \ and\ \bibinfo {author} {\bibfnamefont {R.}~\bibnamefont
  {Easther}},\ }\href {\doibase 10.1103/PhysRevD.103.063525} {\bibfield
  {journal} {\bibinfo  {journal} {Phys. Rev. D}\ }\textbf {\bibinfo {volume}
  {103}},\ \bibinfo {pages} {063525} (\bibinfo {year} {2021})}\BibitemShut
  {NoStop}%
\bibitem [{\citenamefont {Eggemeier}\ \emph {et~al.}(2022)\citenamefont
  {Eggemeier}, \citenamefont {Schwabe}, \citenamefont {Niemeyer},\ and\
  \citenamefont {Easther}}]{Eggemeier2021}%
  \BibitemOpen
  \bibfield  {author} {\bibinfo {author} {\bibfnamefont {B.}~\bibnamefont
  {Eggemeier}}, \bibinfo {author} {\bibfnamefont {B.}~\bibnamefont {Schwabe}},
  \bibinfo {author} {\bibfnamefont {J.~C.}\ \bibnamefont {Niemeyer}}, \ and\
  \bibinfo {author} {\bibfnamefont {R.}~\bibnamefont {Easther}},\ }\href
  {\doibase 10.1103/PhysRevD.105.023516} {\bibfield  {journal} {\bibinfo
  {journal} {Phys. Rev. D}\ }\textbf {\bibinfo {volume} {105}},\ \bibinfo
  {pages} {023516} (\bibinfo {year} {2022})},\ \Eprint
  {http://arxiv.org/abs/2110.15109} {arXiv:2110.15109 [astro-ph.CO]}
  \BibitemShut {NoStop}%
\bibitem [{\citenamefont {Eggemeier}\ \emph
  {et~al.}(2023{\natexlab{a}})\citenamefont {Eggemeier}, \citenamefont
  {Niemeyer}, \citenamefont {Jedamzik},\ and\ \citenamefont
  {Easther}}]{Eggemeier2022}%
  \BibitemOpen
  \bibfield  {author} {\bibinfo {author} {\bibfnamefont {B.}~\bibnamefont
  {Eggemeier}}, \bibinfo {author} {\bibfnamefont {J.~C.}\ \bibnamefont
  {Niemeyer}}, \bibinfo {author} {\bibfnamefont {K.}~\bibnamefont {Jedamzik}},
  \ and\ \bibinfo {author} {\bibfnamefont {R.}~\bibnamefont {Easther}},\ }\href
  {\doibase 10.1103/PhysRevD.107.043503} {\bibfield  {journal} {\bibinfo
  {journal} {Phys. Rev. D}\ }\textbf {\bibinfo {volume} {107}},\ \bibinfo
  {pages} {043503} (\bibinfo {year} {2023}{\natexlab{a}})},\ \Eprint
  {http://arxiv.org/abs/2212.00425} {arXiv:2212.00425 [astro-ph.CO]}
  \BibitemShut {NoStop}%
\bibitem [{\citenamefont {Jedamzik}\ \emph
  {et~al.}(2010{\natexlab{b}})\citenamefont {Jedamzik}, \citenamefont
  {Lemoine},\ and\ \citenamefont {Martin}}]{Jedamzik2010_grav}%
  \BibitemOpen
  \bibfield  {author} {\bibinfo {author} {\bibfnamefont {K.}~\bibnamefont
  {Jedamzik}}, \bibinfo {author} {\bibfnamefont {M.}~\bibnamefont {Lemoine}}, \
  and\ \bibinfo {author} {\bibfnamefont {J.}~\bibnamefont {Martin}},\ }\href
  {\doibase 10.1088/1475-7516/2010/04/021} {\bibfield  {journal} {\bibinfo
  {journal} {JCAP}\ }\textbf {\bibinfo {volume} {04}},\ \bibinfo {pages} {021}
  (\bibinfo {year} {2010}{\natexlab{b}})},\ \Eprint
  {http://arxiv.org/abs/1002.3278} {arXiv:1002.3278 [astro-ph.CO]} \BibitemShut
  {NoStop}%
\bibitem [{\citenamefont {Hu}\ \emph {et~al.}(2000)\citenamefont {Hu},
  \citenamefont {Barkana},\ and\ \citenamefont {Gruzinov}}]{Hu2000}%
  \BibitemOpen
  \bibfield  {author} {\bibinfo {author} {\bibfnamefont {W.}~\bibnamefont
  {Hu}}, \bibinfo {author} {\bibfnamefont {R.}~\bibnamefont {Barkana}}, \ and\
  \bibinfo {author} {\bibfnamefont {A.}~\bibnamefont {Gruzinov}},\ }\href
  {\doibase 10.1103/PhysRevLett.85.1158} {\bibfield  {journal} {\bibinfo
  {journal} {Phys. Rev. Lett.}\ }\textbf {\bibinfo {volume} {85}},\ \bibinfo
  {pages} {1158} (\bibinfo {year} {2000})}\BibitemShut {NoStop}%
\bibitem [{\citenamefont {Hui}\ \emph {et~al.}(2017)\citenamefont {Hui},
  \citenamefont {Ostriker}, \citenamefont {Tremaine},\ and\ \citenamefont
  {Witten}}]{Hui2017}%
  \BibitemOpen
  \bibfield  {author} {\bibinfo {author} {\bibfnamefont {L.}~\bibnamefont
  {Hui}}, \bibinfo {author} {\bibfnamefont {J.~P.}\ \bibnamefont {Ostriker}},
  \bibinfo {author} {\bibfnamefont {S.}~\bibnamefont {Tremaine}}, \ and\
  \bibinfo {author} {\bibfnamefont {E.}~\bibnamefont {Witten}},\ }\href
  {\doibase 10.1103/PhysRevD.95.043541} {\bibfield  {journal} {\bibinfo
  {journal} {Phys. Rev. D}\ }\textbf {\bibinfo {volume} {95}},\ \bibinfo
  {pages} {043541} (\bibinfo {year} {2017})}\BibitemShut {NoStop}%
\bibitem [{\citenamefont {{Tkachev}}(1986)}]{Tkachev1986}%
  \BibitemOpen
  \bibfield  {author} {\bibinfo {author} {\bibfnamefont {I.~I.}\ \bibnamefont
  {{Tkachev}}},\ }\href@noop {} {\bibfield  {journal} {\bibinfo  {journal}
  {Soviet Astronomy Letters}\ }\textbf {\bibinfo {volume} {12}},\ \bibinfo
  {pages} {305} (\bibinfo {year} {1986})}\BibitemShut {NoStop}%
\bibitem [{\citenamefont {Tkachev}(1991)}]{Tkachev1991}%
  \BibitemOpen
  \bibfield  {author} {\bibinfo {author} {\bibfnamefont {I.}~\bibnamefont
  {Tkachev}},\ }\href {\doibase https://doi.org/10.1016/0370-2693(91)90330-S}
  {\bibfield  {journal} {\bibinfo  {journal} {Physics Letters B}\ }\textbf
  {\bibinfo {volume} {261}},\ \bibinfo {pages} {289} (\bibinfo {year}
  {1991})}\BibitemShut {NoStop}%
\bibitem [{\citenamefont {Schive}\ \emph
  {et~al.}(2014{\natexlab{a}})\citenamefont {Schive}, \citenamefont {Chiueh},\
  and\ \citenamefont {Broadhurst}}]{Schive2014_nature}%
  \BibitemOpen
  \bibfield  {author} {\bibinfo {author} {\bibfnamefont {H.-Y.}\ \bibnamefont
  {Schive}}, \bibinfo {author} {\bibfnamefont {T.}~\bibnamefont {Chiueh}}, \
  and\ \bibinfo {author} {\bibfnamefont {T.}~\bibnamefont {Broadhurst}},\
  }\href {\doibase 10.1038/nphys2996} {\bibfield  {journal} {\bibinfo
  {journal} {Nature Phys.}\ }\textbf {\bibinfo {volume} {10}},\ \bibinfo
  {pages} {496} (\bibinfo {year} {2014}{\natexlab{a}})},\ \Eprint
  {http://arxiv.org/abs/1406.6586} {arXiv:1406.6586 [astro-ph.GA]} \BibitemShut
  {NoStop}%
\bibitem [{\citenamefont {Schive}\ \emph
  {et~al.}(2014{\natexlab{b}})\citenamefont {Schive}, \citenamefont {Liao},
  \citenamefont {Woo}, \citenamefont {Wong}, \citenamefont {Chiueh},
  \citenamefont {Broadhurst},\ and\ \citenamefont {Hwang}}]{Schive2014_prl}%
  \BibitemOpen
  \bibfield  {author} {\bibinfo {author} {\bibfnamefont {H.-Y.}\ \bibnamefont
  {Schive}}, \bibinfo {author} {\bibfnamefont {M.-H.}\ \bibnamefont {Liao}},
  \bibinfo {author} {\bibfnamefont {T.-P.}\ \bibnamefont {Woo}}, \bibinfo
  {author} {\bibfnamefont {S.-K.}\ \bibnamefont {Wong}}, \bibinfo {author}
  {\bibfnamefont {T.}~\bibnamefont {Chiueh}}, \bibinfo {author} {\bibfnamefont
  {T.}~\bibnamefont {Broadhurst}}, \ and\ \bibinfo {author} {\bibfnamefont
  {W.-Y.~P.}\ \bibnamefont {Hwang}},\ }\href {\doibase
  10.1103/PhysRevLett.113.261302} {\bibfield  {journal} {\bibinfo  {journal}
  {Phys. Rev. Lett.}\ }\textbf {\bibinfo {volume} {113}},\ \bibinfo {pages}
  {261302} (\bibinfo {year} {2014}{\natexlab{b}})}\BibitemShut {NoStop}%
\bibitem [{\citenamefont {Schwabe}\ \emph {et~al.}(2016)\citenamefont
  {Schwabe}, \citenamefont {Niemeyer},\ and\ \citenamefont
  {Engels}}]{Schwabe2016}%
  \BibitemOpen
  \bibfield  {author} {\bibinfo {author} {\bibfnamefont {B.}~\bibnamefont
  {Schwabe}}, \bibinfo {author} {\bibfnamefont {J.~C.}\ \bibnamefont
  {Niemeyer}}, \ and\ \bibinfo {author} {\bibfnamefont {J.~F.}\ \bibnamefont
  {Engels}},\ }\href {\doibase 10.1103/PhysRevD.94.043513} {\bibfield
  {journal} {\bibinfo  {journal} {Phys. Rev. D}\ }\textbf {\bibinfo {volume}
  {94}},\ \bibinfo {pages} {043513} (\bibinfo {year} {2016})}\BibitemShut
  {NoStop}%
\bibitem [{\citenamefont {Veltmaat}\ \emph {et~al.}(2018)\citenamefont
  {Veltmaat}, \citenamefont {Niemeyer},\ and\ \citenamefont
  {Schwabe}}]{Veltmaat2018}%
  \BibitemOpen
  \bibfield  {author} {\bibinfo {author} {\bibfnamefont {J.}~\bibnamefont
  {Veltmaat}}, \bibinfo {author} {\bibfnamefont {J.~C.}\ \bibnamefont
  {Niemeyer}}, \ and\ \bibinfo {author} {\bibfnamefont {B.}~\bibnamefont
  {Schwabe}},\ }\href {\doibase 10.1103/PhysRevD.98.043509} {\bibfield
  {journal} {\bibinfo  {journal} {Phys. Rev.}\ }\textbf {\bibinfo {volume}
  {D98}},\ \bibinfo {pages} {043509} (\bibinfo {year} {2018})},\ \Eprint
  {http://arxiv.org/abs/1804.09647} {arXiv:1804.09647 [astro-ph.CO]}
  \BibitemShut {NoStop}%
\bibitem [{\citenamefont {Mocz}\ \emph {et~al.}(2017)\citenamefont {Mocz},
  \citenamefont {Vogelsberger}, \citenamefont {Robles}, \citenamefont {Zavala},
  \citenamefont {Boylan-Kolchin}, \citenamefont {Fialkov},\ and\ \citenamefont
  {Hernquist}}]{Mocz2017}%
  \BibitemOpen
  \bibfield  {author} {\bibinfo {author} {\bibfnamefont {P.}~\bibnamefont
  {Mocz}}, \bibinfo {author} {\bibfnamefont {M.}~\bibnamefont {Vogelsberger}},
  \bibinfo {author} {\bibfnamefont {V.~H.}\ \bibnamefont {Robles}}, \bibinfo
  {author} {\bibfnamefont {J.}~\bibnamefont {Zavala}}, \bibinfo {author}
  {\bibfnamefont {M.}~\bibnamefont {Boylan-Kolchin}}, \bibinfo {author}
  {\bibfnamefont {A.}~\bibnamefont {Fialkov}}, \ and\ \bibinfo {author}
  {\bibfnamefont {L.}~\bibnamefont {Hernquist}},\ }\href {\doibase
  10.1093/mnras/stx1887} {\bibfield  {journal} {\bibinfo  {journal} {Monthly
  Notices of the Royal Astronomical Society}\ }\textbf {\bibinfo {volume}
  {471}},\ \bibinfo {pages} {4559} (\bibinfo {year} {2017})},\ \Eprint
  {http://arxiv.org/abs/https://academic.oup.com/mnras/article-pdf/471/4/4559/19609125/stx1887.pdf}
  {https://academic.oup.com/mnras/article-pdf/471/4/4559/19609125/stx1887.pdf}
  \BibitemShut {NoStop}%
\bibitem [{\citenamefont {Mocz}\ \emph {et~al.}(2018)\citenamefont {Mocz},
  \citenamefont {Lancaster}, \citenamefont {Fialkov}, \citenamefont {Becerra},\
  and\ \citenamefont {Chavanis}}]{Mocz2018}%
  \BibitemOpen
  \bibfield  {author} {\bibinfo {author} {\bibfnamefont {P.}~\bibnamefont
  {Mocz}}, \bibinfo {author} {\bibfnamefont {L.}~\bibnamefont {Lancaster}},
  \bibinfo {author} {\bibfnamefont {A.}~\bibnamefont {Fialkov}}, \bibinfo
  {author} {\bibfnamefont {F.}~\bibnamefont {Becerra}}, \ and\ \bibinfo
  {author} {\bibfnamefont {P.-H.}\ \bibnamefont {Chavanis}},\ }\href {\doibase
  10.1103/PhysRevD.97.083519} {\bibfield  {journal} {\bibinfo  {journal} {Phys.
  Rev. D}\ }\textbf {\bibinfo {volume} {97}},\ \bibinfo {pages} {083519}
  (\bibinfo {year} {2018})}\BibitemShut {NoStop}%
\bibitem [{\citenamefont {Levkov}\ \emph {et~al.}(2018)\citenamefont {Levkov},
  \citenamefont {Panin},\ and\ \citenamefont {Tkachev}}]{Levkov2018}%
  \BibitemOpen
  \bibfield  {author} {\bibinfo {author} {\bibfnamefont {D.~G.}\ \bibnamefont
  {Levkov}}, \bibinfo {author} {\bibfnamefont {A.~G.}\ \bibnamefont {Panin}}, \
  and\ \bibinfo {author} {\bibfnamefont {I.~I.}\ \bibnamefont {Tkachev}},\
  }\href {\doibase 10.1103/PhysRevLett.121.151301} {\bibfield  {journal}
  {\bibinfo  {journal} {Phys. Rev. Lett.}\ }\textbf {\bibinfo {volume} {121}},\
  \bibinfo {pages} {151301} (\bibinfo {year} {2018})}\BibitemShut {NoStop}%
\bibitem [{\citenamefont {Edwards}\ \emph {et~al.}(2018)\citenamefont
  {Edwards}, \citenamefont {Kendall}, \citenamefont {Hotchkiss},\ and\
  \citenamefont {Easther}}]{Pyultralight}%
  \BibitemOpen
  \bibfield  {author} {\bibinfo {author} {\bibfnamefont {F.}~\bibnamefont
  {Edwards}}, \bibinfo {author} {\bibfnamefont {E.}~\bibnamefont {Kendall}},
  \bibinfo {author} {\bibfnamefont {S.}~\bibnamefont {Hotchkiss}}, \ and\
  \bibinfo {author} {\bibfnamefont {R.}~\bibnamefont {Easther}},\ }\href
  {\doibase 10.1088/1475-7516/2018/10/027} {\bibfield  {journal} {\bibinfo
  {journal} {JCAP}\ }\textbf {\bibinfo {volume} {10}},\ \bibinfo {pages} {027}
  (\bibinfo {year} {2018})},\ \Eprint {http://arxiv.org/abs/1807.04037}
  {arXiv:1807.04037 [astro-ph.CO]} \BibitemShut {NoStop}%
\bibitem [{\citenamefont {{Eggemeier}}\ and\ \citenamefont
  {{Niemeyer}}(2019)}]{Eggemeier2019PRD}%
  \BibitemOpen
  \bibfield  {author} {\bibinfo {author} {\bibfnamefont {B.}~\bibnamefont
  {{Eggemeier}}}\ and\ \bibinfo {author} {\bibfnamefont {J.~C.}\ \bibnamefont
  {{Niemeyer}}},\ }\href {\doibase 10.1103/PhysRevD.100.063528} {\bibfield
  {journal} {\bibinfo  {journal} {\prd}\ }\textbf {\bibinfo {volume} {100}},\
  \bibinfo {eid} {063528} (\bibinfo {year} {2019})},\ \Eprint
  {http://arxiv.org/abs/1906.01348} {arXiv:1906.01348 [astro-ph.CO]}
  \BibitemShut {NoStop}%
\bibitem [{\citenamefont {Chen}\ \emph {et~al.}(2020)\citenamefont {Chen},
  \citenamefont {Du}, \citenamefont {Lentz}, \citenamefont {Marsh},\ and\
  \citenamefont {Niemeyer}}]{Jiajun2020}%
  \BibitemOpen
  \bibfield  {author} {\bibinfo {author} {\bibfnamefont {J.}~\bibnamefont
  {Chen}}, \bibinfo {author} {\bibfnamefont {X.}~\bibnamefont {Du}}, \bibinfo
  {author} {\bibfnamefont {E.~W.}\ \bibnamefont {Lentz}}, \bibinfo {author}
  {\bibfnamefont {D.~J.~E.}\ \bibnamefont {Marsh}}, \ and\ \bibinfo {author}
  {\bibfnamefont {J.~C.}\ \bibnamefont {Niemeyer}},\ }\href@noop {} {\
  (\bibinfo {year} {2020})},\ \Eprint {http://arxiv.org/abs/2011.01333}
  {arXiv:2011.01333 [astro-ph.CO]} \BibitemShut {NoStop}%
\bibitem [{\citenamefont {Schwabe}\ \emph {et~al.}(2020)\citenamefont
  {Schwabe}, \citenamefont {Gosenca}, \citenamefont {Behrens}, \citenamefont
  {Niemeyer},\ and\ \citenamefont {Easther}}]{Schwabe2020}%
  \BibitemOpen
  \bibfield  {author} {\bibinfo {author} {\bibfnamefont {B.}~\bibnamefont
  {Schwabe}}, \bibinfo {author} {\bibfnamefont {M.}~\bibnamefont {Gosenca}},
  \bibinfo {author} {\bibfnamefont {C.}~\bibnamefont {Behrens}}, \bibinfo
  {author} {\bibfnamefont {J.~C.}\ \bibnamefont {Niemeyer}}, \ and\ \bibinfo
  {author} {\bibfnamefont {R.}~\bibnamefont {Easther}},\ }\href {\doibase
  10.1103/PhysRevD.102.083518} {\bibfield  {journal} {\bibinfo  {journal}
  {Phys. Rev. D}\ }\textbf {\bibinfo {volume} {102}},\ \bibinfo {pages}
  {083518} (\bibinfo {year} {2020})}\BibitemShut {NoStop}%
\bibitem [{\citenamefont {Schwabe}\ and\ \citenamefont
  {Niemeyer}(2022)}]{Schwabe2021}%
  \BibitemOpen
  \bibfield  {author} {\bibinfo {author} {\bibfnamefont {B.}~\bibnamefont
  {Schwabe}}\ and\ \bibinfo {author} {\bibfnamefont {J.~C.}\ \bibnamefont
  {Niemeyer}},\ }\href {\doibase 10.1103/PhysRevLett.128.181301} {\bibfield
  {journal} {\bibinfo  {journal} {Phys. Rev. Lett.}\ }\textbf {\bibinfo
  {volume} {128}},\ \bibinfo {pages} {181301} (\bibinfo {year} {2022})},\
  \Eprint {http://arxiv.org/abs/2110.09145} {arXiv:2110.09145 [astro-ph.CO]}
  \BibitemShut {NoStop}%
\bibitem [{\citenamefont {Chan}\ \emph {et~al.}(2022)\citenamefont {Chan},
  \citenamefont {Sibiryakov},\ and\ \citenamefont {Xue}}]{Chan2022}%
  \BibitemOpen
  \bibfield  {author} {\bibinfo {author} {\bibfnamefont {J.~H.-H.}\
  \bibnamefont {Chan}}, \bibinfo {author} {\bibfnamefont {S.}~\bibnamefont
  {Sibiryakov}}, \ and\ \bibinfo {author} {\bibfnamefont {W.}~\bibnamefont
  {Xue}},\ }\href@noop {} {\  (\bibinfo {year} {2022})},\ \Eprint
  {http://arxiv.org/abs/2207.04057} {arXiv:2207.04057 [astro-ph.CO]}
  \BibitemShut {NoStop}%
\bibitem [{\citenamefont {Gosenca}\ \emph {et~al.}(2023)\citenamefont
  {Gosenca}, \citenamefont {Eberhardt}, \citenamefont {Wang}, \citenamefont
  {Eggemeier}, \citenamefont {Kendall}, \citenamefont {Zagorac},\ and\
  \citenamefont {Easther}}]{Gosenca2023}%
  \BibitemOpen
  \bibfield  {author} {\bibinfo {author} {\bibfnamefont {M.}~\bibnamefont
  {Gosenca}}, \bibinfo {author} {\bibfnamefont {A.}~\bibnamefont {Eberhardt}},
  \bibinfo {author} {\bibfnamefont {Y.}~\bibnamefont {Wang}}, \bibinfo {author}
  {\bibfnamefont {B.}~\bibnamefont {Eggemeier}}, \bibinfo {author}
  {\bibfnamefont {E.}~\bibnamefont {Kendall}}, \bibinfo {author} {\bibfnamefont
  {J.~L.}\ \bibnamefont {Zagorac}}, \ and\ \bibinfo {author} {\bibfnamefont
  {R.}~\bibnamefont {Easther}},\ }\href {\doibase 10.1103/PhysRevD.107.083014}
  {\bibfield  {journal} {\bibinfo  {journal} {Phys. Rev. D}\ }\textbf {\bibinfo
  {volume} {107}},\ \bibinfo {pages} {083014} (\bibinfo {year} {2023})},\
  \Eprint {http://arxiv.org/abs/2301.07114} {arXiv:2301.07114 [astro-ph.CO]}
  \BibitemShut {NoStop}%
\bibitem [{\citenamefont {Dmitriev}\ \emph {et~al.}(2023)\citenamefont
  {Dmitriev}, \citenamefont {Levkov}, \citenamefont {Panin},\ and\
  \citenamefont {Tkachev}}]{Dmitriev2023}%
  \BibitemOpen
  \bibfield  {author} {\bibinfo {author} {\bibfnamefont {A.~S.}\ \bibnamefont
  {Dmitriev}}, \bibinfo {author} {\bibfnamefont {D.~G.}\ \bibnamefont
  {Levkov}}, \bibinfo {author} {\bibfnamefont {A.~G.}\ \bibnamefont {Panin}}, \
  and\ \bibinfo {author} {\bibfnamefont {I.~I.}\ \bibnamefont {Tkachev}},\
  }\href@noop {} {\  (\bibinfo {year} {2023})},\ \Eprint
  {http://arxiv.org/abs/2305.01005} {arXiv:2305.01005 [astro-ph.CO]}
  \BibitemShut {NoStop}%
\bibitem [{\citenamefont {{Widrow}}\ and\ \citenamefont
  {{Kaiser}}(1993)}]{Widrow1993}%
  \BibitemOpen
  \bibfield  {author} {\bibinfo {author} {\bibfnamefont {L.~M.}\ \bibnamefont
  {{Widrow}}}\ and\ \bibinfo {author} {\bibfnamefont {N.}~\bibnamefont
  {{Kaiser}}},\ }\href {\doibase 10.1086/187073} {\bibfield  {journal}
  {\bibinfo  {journal} {\apjl}\ }\textbf {\bibinfo {volume} {416}},\ \bibinfo
  {pages} {L71} (\bibinfo {year} {1993})}\BibitemShut {NoStop}%
\bibitem [{\citenamefont {Uhlemann}\ \emph {et~al.}(2014)\citenamefont
  {Uhlemann}, \citenamefont {Kopp},\ and\ \citenamefont
  {Haugg}}]{Uhlemann2014}%
  \BibitemOpen
  \bibfield  {author} {\bibinfo {author} {\bibfnamefont {C.}~\bibnamefont
  {Uhlemann}}, \bibinfo {author} {\bibfnamefont {M.}~\bibnamefont {Kopp}}, \
  and\ \bibinfo {author} {\bibfnamefont {T.}~\bibnamefont {Haugg}},\ }\href
  {\doibase 10.1103/PhysRevD.90.023517} {\bibfield  {journal} {\bibinfo
  {journal} {Phys. Rev. D}\ }\textbf {\bibinfo {volume} {90}},\ \bibinfo
  {pages} {023517} (\bibinfo {year} {2014})}\BibitemShut {NoStop}%
\bibitem [{\citenamefont {{Padilla}}\ \emph {et~al.}(2022)\citenamefont
  {{Padilla}}, \citenamefont {{Hidalgo}},\ and\ \citenamefont
  {{Malik}}}]{Padilla2022}%
  \BibitemOpen
  \bibfield  {author} {\bibinfo {author} {\bibfnamefont {L.~E.}\ \bibnamefont
  {{Padilla}}}, \bibinfo {author} {\bibfnamefont {J.~C.}\ \bibnamefont
  {{Hidalgo}}}, \ and\ \bibinfo {author} {\bibfnamefont {K.~A.}\ \bibnamefont
  {{Malik}}},\ }\href {\doibase 10.1103/PhysRevD.106.023519} {\bibfield
  {journal} {\bibinfo  {journal} {\prd}\ }\textbf {\bibinfo {volume} {106}},\
  \bibinfo {eid} {023519} (\bibinfo {year} {2022})},\ \Eprint
  {http://arxiv.org/abs/2110.14584} {arXiv:2110.14584 [astro-ph.CO]}
  \BibitemShut {NoStop}%
\bibitem [{\citenamefont {Hidalgo}\ \emph {et~al.}(2023)\citenamefont
  {Hidalgo}, \citenamefont {Padilla},\ and\ \citenamefont
  {German}}]{Hidalgo2022}%
  \BibitemOpen
  \bibfield  {author} {\bibinfo {author} {\bibfnamefont {J.~C.}\ \bibnamefont
  {Hidalgo}}, \bibinfo {author} {\bibfnamefont {L.~E.}\ \bibnamefont
  {Padilla}}, \ and\ \bibinfo {author} {\bibfnamefont {G.}~\bibnamefont
  {German}},\ }\href {\doibase 10.1103/PhysRevD.107.063519} {\bibfield
  {journal} {\bibinfo  {journal} {Phys. Rev. D}\ }\textbf {\bibinfo {volume}
  {107}},\ \bibinfo {pages} {063519} (\bibinfo {year} {2023})},\ \Eprint
  {http://arxiv.org/abs/2208.09462} {arXiv:2208.09462 [astro-ph.CO]}
  \BibitemShut {NoStop}%
\bibitem [{\citenamefont {Punturo}\ \emph {et~al.}(2010)\citenamefont {Punturo}
  \emph {et~al.}}]{ET}%
  \BibitemOpen
  \bibfield  {author} {\bibinfo {author} {\bibfnamefont {M.}~\bibnamefont
  {Punturo}} \emph {et~al.},\ }\href {\doibase 10.1088/0264-9381/27/19/194002}
  {\bibfield  {journal} {\bibinfo  {journal} {Class. Quant. Grav.}\ }\textbf
  {\bibinfo {volume} {27}},\ \bibinfo {pages} {194002} (\bibinfo {year}
  {2010})}\BibitemShut {NoStop}%
\bibitem [{\citenamefont {Cutler}\ and\ \citenamefont {Holz}(2009)}]{BBO}%
  \BibitemOpen
  \bibfield  {author} {\bibinfo {author} {\bibfnamefont {C.}~\bibnamefont
  {Cutler}}\ and\ \bibinfo {author} {\bibfnamefont {D.~E.}\ \bibnamefont
  {Holz}},\ }\href {\doibase 10.1103/PhysRevD.80.104009} {\bibfield  {journal}
  {\bibinfo  {journal} {Phys. Rev. D}\ }\textbf {\bibinfo {volume} {80}},\
  \bibinfo {pages} {104009} (\bibinfo {year} {2009})}\BibitemShut {NoStop}%
\bibitem [{\citenamefont {Silverstein}\ and\ \citenamefont
  {Westphal}(2008)}]{Silverstein2008}%
  \BibitemOpen
  \bibfield  {author} {\bibinfo {author} {\bibfnamefont {E.}~\bibnamefont
  {Silverstein}}\ and\ \bibinfo {author} {\bibfnamefont {A.}~\bibnamefont
  {Westphal}},\ }\href {\doibase 10.1103/PhysRevD.78.106003} {\bibfield
  {journal} {\bibinfo  {journal} {Phys. Rev. D}\ }\textbf {\bibinfo {volume}
  {78}},\ \bibinfo {pages} {106003} (\bibinfo {year} {2008})},\ \Eprint
  {http://arxiv.org/abs/0803.3085} {arXiv:0803.3085 [hep-th]} \BibitemShut
  {NoStop}%
\bibitem [{\citenamefont {McAllister}\ \emph {et~al.}(2010)\citenamefont
  {McAllister}, \citenamefont {Silverstein},\ and\ \citenamefont
  {Westphal}}]{McAllister2008}%
  \BibitemOpen
  \bibfield  {author} {\bibinfo {author} {\bibfnamefont {L.}~\bibnamefont
  {McAllister}}, \bibinfo {author} {\bibfnamefont {E.}~\bibnamefont
  {Silverstein}}, \ and\ \bibinfo {author} {\bibfnamefont {A.}~\bibnamefont
  {Westphal}},\ }\href {\doibase 10.1103/PhysRevD.82.046003} {\bibfield
  {journal} {\bibinfo  {journal} {Phys. Rev. D}\ }\textbf {\bibinfo {volume}
  {82}},\ \bibinfo {pages} {046003} (\bibinfo {year} {2010})},\ \Eprint
  {http://arxiv.org/abs/0808.0706} {arXiv:0808.0706 [hep-th]} \BibitemShut
  {NoStop}%
\bibitem [{\citenamefont {Amin}\ and\ \citenamefont
  {Mocz}(2019)}]{Amin:2019ums}%
  \BibitemOpen
  \bibfield  {author} {\bibinfo {author} {\bibfnamefont {M.~A.}\ \bibnamefont
  {Amin}}\ and\ \bibinfo {author} {\bibfnamefont {P.}~\bibnamefont {Mocz}},\
  }\href {\doibase 10.1103/PhysRevD.100.063507} {\bibfield  {journal} {\bibinfo
   {journal} {Phys. Rev. D}\ }\textbf {\bibinfo {volume} {100}},\ \bibinfo
  {pages} {063507} (\bibinfo {year} {2019})},\ \Eprint
  {http://arxiv.org/abs/1902.07261} {arXiv:1902.07261 [astro-ph.CO]}
  \BibitemShut {NoStop}%
\bibitem [{\citenamefont {Amin}\ and\ \citenamefont
  {Shirokoff}(2010)}]{Amin:2010jq}%
  \BibitemOpen
  \bibfield  {author} {\bibinfo {author} {\bibfnamefont {M.~A.}\ \bibnamefont
  {Amin}}\ and\ \bibinfo {author} {\bibfnamefont {D.}~\bibnamefont
  {Shirokoff}},\ }\href {\doibase 10.1103/PhysRevD.81.085045} {\bibfield
  {journal} {\bibinfo  {journal} {Phys. Rev. D}\ }\textbf {\bibinfo {volume}
  {81}},\ \bibinfo {pages} {085045} (\bibinfo {year} {2010})},\ \Eprint
  {http://arxiv.org/abs/1002.3380} {arXiv:1002.3380 [astro-ph.CO]} \BibitemShut
  {NoStop}%
\bibitem [{\citenamefont {van Dissel}\ \emph {et~al.}(2023)\citenamefont {van
  Dissel}, \citenamefont {Pujolas},\ and\ \citenamefont
  {Sfakianakis}}]{vanDissel:2023zva}%
  \BibitemOpen
  \bibfield  {author} {\bibinfo {author} {\bibfnamefont {F.}~\bibnamefont {van
  Dissel}}, \bibinfo {author} {\bibfnamefont {O.}~\bibnamefont {Pujolas}}, \
  and\ \bibinfo {author} {\bibfnamefont {E.~I.}\ \bibnamefont {Sfakianakis}},\
  }\href {\doibase 10.1007/JHEP07(2023)194} {\bibfield  {journal} {\bibinfo
  {journal} {JHEP}\ }\textbf {\bibinfo {volume} {07}},\ \bibinfo {pages} {194}
  (\bibinfo {year} {2023})},\ \Eprint {http://arxiv.org/abs/2303.16072}
  {arXiv:2303.16072 [hep-th]} \BibitemShut {NoStop}%
\bibitem [{\citenamefont {Amin}\ \emph {et~al.}(2014)\citenamefont {Amin},
  \citenamefont {Hertzberg}, \citenamefont {Kaiser},\ and\ \citenamefont
  {Karouby}}]{Amin2014_review}%
  \BibitemOpen
  \bibfield  {author} {\bibinfo {author} {\bibfnamefont {M.~A.}\ \bibnamefont
  {Amin}}, \bibinfo {author} {\bibfnamefont {M.~P.}\ \bibnamefont {Hertzberg}},
  \bibinfo {author} {\bibfnamefont {D.~I.}\ \bibnamefont {Kaiser}}, \ and\
  \bibinfo {author} {\bibfnamefont {J.}~\bibnamefont {Karouby}},\ }\href
  {\doibase 10.1142/S0218271815300037} {\bibfield  {journal} {\bibinfo
  {journal} {Int. J. Mod. Phys. D}\ }\textbf {\bibinfo {volume} {24}},\
  \bibinfo {pages} {1530003} (\bibinfo {year} {2014})},\ \Eprint
  {http://arxiv.org/abs/1410.3808} {arXiv:1410.3808 [hep-ph]} \BibitemShut
  {NoStop}%
\bibitem [{\citenamefont {Ruffini}\ and\ \citenamefont
  {Bonazzola}(1969)}]{Ruffini1969}%
  \BibitemOpen
  \bibfield  {author} {\bibinfo {author} {\bibfnamefont {R.}~\bibnamefont
  {Ruffini}}\ and\ \bibinfo {author} {\bibfnamefont {S.}~\bibnamefont
  {Bonazzola}},\ }\href {\doibase 10.1103/PhysRev.187.1767} {\bibfield
  {journal} {\bibinfo  {journal} {Phys. Rev.}\ }\textbf {\bibinfo {volume}
  {187}},\ \bibinfo {pages} {1767} (\bibinfo {year} {1969})}\BibitemShut
  {NoStop}%
\bibitem [{\citenamefont {Nambu}\ and\ \citenamefont
  {Sasaki}(1990)}]{Nambu1990QuantumPerturbations}%
  \BibitemOpen
  \bibfield  {author} {\bibinfo {author} {\bibfnamefont {Y.}~\bibnamefont
  {Nambu}}\ and\ \bibinfo {author} {\bibfnamefont {M.}~\bibnamefont {Sasaki}},\
  }\href {\doibase 10.1103/PhysRevD.42.3918} {\bibfield  {journal} {\bibinfo
  {journal} {Physical Review D}\ }\textbf {\bibinfo {volume} {42}},\ \bibinfo
  {pages} {3918} (\bibinfo {year} {1990})}\BibitemShut {NoStop}%
\bibitem [{\citenamefont {Adshead}\ \emph {et~al.}(2011)\citenamefont
  {Adshead}, \citenamefont {Easther}, \citenamefont {Pritchard},\ and\
  \citenamefont {Loeb}}]{Adshead2011}%
  \BibitemOpen
  \bibfield  {author} {\bibinfo {author} {\bibfnamefont {P.}~\bibnamefont
  {Adshead}}, \bibinfo {author} {\bibfnamefont {R.}~\bibnamefont {Easther}},
  \bibinfo {author} {\bibfnamefont {J.}~\bibnamefont {Pritchard}}, \ and\
  \bibinfo {author} {\bibfnamefont {A.}~\bibnamefont {Loeb}},\ }\href {\doibase
  10.1088/1475-7516/2011/02/021} {\bibfield  {journal} {\bibinfo  {journal}
  {JCAP}\ }\textbf {\bibinfo {volume} {02}},\ \bibinfo {pages} {021} (\bibinfo
  {year} {2011})},\ \Eprint {http://arxiv.org/abs/1007.3748} {arXiv:1007.3748
  [astro-ph.CO]} \BibitemShut {NoStop}%
\bibitem [{\citenamefont {Felder}\ and\ \citenamefont
  {Tkachev}(2008)}]{Latticeeasy}%
  \BibitemOpen
  \bibfield  {author} {\bibinfo {author} {\bibfnamefont {G.~N.}\ \bibnamefont
  {Felder}}\ and\ \bibinfo {author} {\bibfnamefont {I.}~\bibnamefont
  {Tkachev}},\ }\href {\doibase 10.1016/j.cpc.2008.02.009} {\bibfield
  {journal} {\bibinfo  {journal} {Comput. Phys. Commun.}\ }\textbf {\bibinfo
  {volume} {178}},\ \bibinfo {pages} {929} (\bibinfo {year} {2008})},\ \Eprint
  {http://arxiv.org/abs/hep-ph/0011159} {arXiv:hep-ph/0011159} \BibitemShut
  {NoStop}%
\bibitem [{\citenamefont {{Zhang}}\ \emph {et~al.}(2020)\citenamefont
  {{Zhang}}, \citenamefont {{Myers}}, \citenamefont {{Gott}}, \citenamefont
  {{Almgren}},\ and\ \citenamefont {{Bell}}}]{Amrex}%
  \BibitemOpen
  \bibfield  {author} {\bibinfo {author} {\bibfnamefont {W.}~\bibnamefont
  {{Zhang}}}, \bibinfo {author} {\bibfnamefont {A.}~\bibnamefont {{Myers}}},
  \bibinfo {author} {\bibfnamefont {K.}~\bibnamefont {{Gott}}}, \bibinfo
  {author} {\bibfnamefont {A.}~\bibnamefont {{Almgren}}}, \ and\ \bibinfo
  {author} {\bibfnamefont {J.}~\bibnamefont {{Bell}}},\ }\href {\doibase
  10.48550/arXiv.2009.12009} {\  (\bibinfo {year} {2020}),\
  10.48550/arXiv.2009.12009},\ \Eprint {http://arxiv.org/abs/2009.12009}
  {2009.12009 [cs.MS]} \BibitemShut {NoStop}%
\bibitem [{\citenamefont {Salehian}\ \emph {et~al.}(2020)\citenamefont
  {Salehian}, \citenamefont {Namjoo},\ and\ \citenamefont
  {Kaiser}}]{Salehian2020}%
  \BibitemOpen
  \bibfield  {author} {\bibinfo {author} {\bibfnamefont {B.}~\bibnamefont
  {Salehian}}, \bibinfo {author} {\bibfnamefont {M.~H.}\ \bibnamefont
  {Namjoo}}, \ and\ \bibinfo {author} {\bibfnamefont {D.~I.}\ \bibnamefont
  {Kaiser}},\ }\href {\doibase 10.1007/JHEP07(2020)059} {\bibfield  {journal}
  {\bibinfo  {journal} {JHEP}\ }\textbf {\bibinfo {volume} {07}},\ \bibinfo
  {pages} {059} (\bibinfo {year} {2020})},\ \Eprint
  {http://arxiv.org/abs/2005.05388} {arXiv:2005.05388 [astro-ph.CO]}
  \BibitemShut {NoStop}%
\bibitem [{\citenamefont {Mocz}\ \emph {et~al.}(2020)\citenamefont {Mocz} \emph
  {et~al.}}]{Mocz2019}%
  \BibitemOpen
  \bibfield  {author} {\bibinfo {author} {\bibfnamefont {P.}~\bibnamefont
  {Mocz}} \emph {et~al.},\ }\href {\doibase 10.1093/mnras/staa738} {\bibfield
  {journal} {\bibinfo  {journal} {Mon. Not. Roy. Astron. Soc.}\ }\textbf
  {\bibinfo {volume} {494}},\ \bibinfo {pages} {2027} (\bibinfo {year}
  {2020})},\ \Eprint {http://arxiv.org/abs/1911.05746} {arXiv:1911.05746
  [astro-ph.CO]} \BibitemShut {NoStop}%
\bibitem [{\citenamefont {May}\ and\ \citenamefont {Springel}(2021)}]{May2021}%
  \BibitemOpen
  \bibfield  {author} {\bibinfo {author} {\bibfnamefont {S.}~\bibnamefont
  {May}}\ and\ \bibinfo {author} {\bibfnamefont {V.}~\bibnamefont {Springel}},\
  }\href {\doibase 10.1093/mnras/stab1764} {\bibfield  {journal} {\bibinfo
  {journal} {Mon. Not. Roy. Astron. Soc.}\ }\textbf {\bibinfo {volume} {506}},\
  \bibinfo {pages} {2603} (\bibinfo {year} {2021})},\ \Eprint
  {http://arxiv.org/abs/2101.01828} {arXiv:2101.01828 [astro-ph.CO]}
  \BibitemShut {NoStop}%
\bibitem [{\citenamefont {May}\ and\ \citenamefont {Springel}(2023)}]{May2022}%
  \BibitemOpen
  \bibfield  {author} {\bibinfo {author} {\bibfnamefont {S.}~\bibnamefont
  {May}}\ and\ \bibinfo {author} {\bibfnamefont {V.}~\bibnamefont {Springel}},\
  }\href {\doibase 10.1093/mnras/stad2031} {\bibfield  {journal} {\bibinfo
  {journal} {Mon. Not. Roy. Astron. Soc.}\ }\textbf {\bibinfo {volume} {524}},\
  \bibinfo {pages} {4256} (\bibinfo {year} {2023})},\ \Eprint
  {http://arxiv.org/abs/2209.14886} {arXiv:2209.14886 [astro-ph.CO]}
  \BibitemShut {NoStop}%
\bibitem [{\citenamefont {Eisenstein}\ and\ \citenamefont {Hut}(1998)}]{HOP}%
  \BibitemOpen
  \bibfield  {author} {\bibinfo {author} {\bibfnamefont {D.~J.}\ \bibnamefont
  {Eisenstein}}\ and\ \bibinfo {author} {\bibfnamefont {P.}~\bibnamefont
  {Hut}},\ }\href {\doibase 10.1086/305535} {\bibfield  {journal} {\bibinfo
  {journal} {Astrophys. J.}\ }\textbf {\bibinfo {volume} {498}},\ \bibinfo
  {pages} {137} (\bibinfo {year} {1998})},\ \Eprint
  {http://arxiv.org/abs/astro-ph/9712200} {arXiv:astro-ph/9712200} \BibitemShut
  {NoStop}%
\bibitem [{\citenamefont {Zagorac}\ \emph {et~al.}(2023)\citenamefont
  {Zagorac}, \citenamefont {Kendall}, \citenamefont {Padmanabhan},\ and\
  \citenamefont {Easther}}]{Zagorac2022}%
  \BibitemOpen
  \bibfield  {author} {\bibinfo {author} {\bibfnamefont {J.~L.}\ \bibnamefont
  {Zagorac}}, \bibinfo {author} {\bibfnamefont {E.}~\bibnamefont {Kendall}},
  \bibinfo {author} {\bibfnamefont {N.}~\bibnamefont {Padmanabhan}}, \ and\
  \bibinfo {author} {\bibfnamefont {R.}~\bibnamefont {Easther}},\ }\href
  {\doibase 10.1103/PhysRevD.107.083513} {\bibfield  {journal} {\bibinfo
  {journal} {Phys. Rev. D}\ }\textbf {\bibinfo {volume} {107}},\ \bibinfo
  {pages} {083513} (\bibinfo {year} {2023})},\ \Eprint
  {http://arxiv.org/abs/2212.09349} {arXiv:2212.09349 [astro-ph.CO]}
  \BibitemShut {NoStop}%
\bibitem [{\citenamefont {Kendall}\ \emph {et~al.}(2023)\citenamefont
  {Kendall}, \citenamefont {Gosenca},\ and\ \citenamefont
  {Easther}}]{Kendall2023}%
  \BibitemOpen
  \bibfield  {author} {\bibinfo {author} {\bibfnamefont {E.}~\bibnamefont
  {Kendall}}, \bibinfo {author} {\bibfnamefont {M.}~\bibnamefont {Gosenca}}, \
  and\ \bibinfo {author} {\bibfnamefont {R.}~\bibnamefont {Easther}},\
  }\href@noop {} {\  (\bibinfo {year} {2023})},\ \Eprint
  {http://arxiv.org/abs/2305.10340} {arXiv:2305.10340 [astro-ph.CO]}
  \BibitemShut {NoStop}%
\bibitem [{\citenamefont {Vaquero}\ \emph {et~al.}(2019)\citenamefont
  {Vaquero}, \citenamefont {Redondo},\ and\ \citenamefont
  {Stadler}}]{Vaquero2018}%
  \BibitemOpen
  \bibfield  {author} {\bibinfo {author} {\bibfnamefont {A.}~\bibnamefont
  {Vaquero}}, \bibinfo {author} {\bibfnamefont {J.}~\bibnamefont {Redondo}}, \
  and\ \bibinfo {author} {\bibfnamefont {J.}~\bibnamefont {Stadler}},\ }\href
  {\doibase 10.1088/1475-7516/2019/04/012} {\bibfield  {journal} {\bibinfo
  {journal} {JCAP}\ }\textbf {\bibinfo {volume} {04}},\ \bibinfo {pages} {012}
  (\bibinfo {year} {2019})},\ \Eprint {http://arxiv.org/abs/1809.09241}
  {arXiv:1809.09241 [astro-ph.CO]} \BibitemShut {NoStop}%
\bibitem [{\citenamefont {Buschmann}\ \emph {et~al.}(2020)\citenamefont
  {Buschmann}, \citenamefont {Foster},\ and\ \citenamefont
  {Safdi}}]{Buschmann2019}%
  \BibitemOpen
  \bibfield  {author} {\bibinfo {author} {\bibfnamefont {M.}~\bibnamefont
  {Buschmann}}, \bibinfo {author} {\bibfnamefont {J.~W.}\ \bibnamefont
  {Foster}}, \ and\ \bibinfo {author} {\bibfnamefont {B.~R.}\ \bibnamefont
  {Safdi}},\ }\href {\doibase 10.1103/PhysRevLett.124.161103} {\bibfield
  {journal} {\bibinfo  {journal} {Phys. Rev. Lett.}\ }\textbf {\bibinfo
  {volume} {124}},\ \bibinfo {pages} {161103} (\bibinfo {year} {2020})},\
  \Eprint {http://arxiv.org/abs/1906.00967} {arXiv:1906.00967 [astro-ph.CO]}
  \BibitemShut {NoStop}%
\bibitem [{\citenamefont {O'Hare}\ \emph {et~al.}(2022)\citenamefont {O'Hare},
  \citenamefont {Pierobon}, \citenamefont {Redondo},\ and\ \citenamefont
  {Wong}}]{OHare2021}%
  \BibitemOpen
  \bibfield  {author} {\bibinfo {author} {\bibfnamefont {C.~A.~J.}\
  \bibnamefont {O'Hare}}, \bibinfo {author} {\bibfnamefont {G.}~\bibnamefont
  {Pierobon}}, \bibinfo {author} {\bibfnamefont {J.}~\bibnamefont {Redondo}}, \
  and\ \bibinfo {author} {\bibfnamefont {Y.~Y.~Y.}\ \bibnamefont {Wong}},\
  }\href {\doibase 10.1103/PhysRevD.105.055025} {\bibfield  {journal} {\bibinfo
   {journal} {Phys. Rev. D}\ }\textbf {\bibinfo {volume} {105}},\ \bibinfo
  {pages} {055025} (\bibinfo {year} {2022})},\ \Eprint
  {http://arxiv.org/abs/2112.05117} {arXiv:2112.05117 [hep-ph]} \BibitemShut
  {NoStop}%
\bibitem [{\citenamefont {Eggemeier}\ \emph
  {et~al.}(2023{\natexlab{b}})\citenamefont {Eggemeier}, \citenamefont
  {O'Hare}, \citenamefont {Pierobon}, \citenamefont {Redondo},\ and\
  \citenamefont {Wong}}]{Eggemeier2022_voids}%
  \BibitemOpen
  \bibfield  {author} {\bibinfo {author} {\bibfnamefont {B.}~\bibnamefont
  {Eggemeier}}, \bibinfo {author} {\bibfnamefont {C.~A.~J.}\ \bibnamefont
  {O'Hare}}, \bibinfo {author} {\bibfnamefont {G.}~\bibnamefont {Pierobon}},
  \bibinfo {author} {\bibfnamefont {J.}~\bibnamefont {Redondo}}, \ and\
  \bibinfo {author} {\bibfnamefont {Y.~Y.~Y.}\ \bibnamefont {Wong}},\ }\href
  {\doibase 10.1103/PhysRevD.107.083510} {\bibfield  {journal} {\bibinfo
  {journal} {Phys. Rev. D}\ }\textbf {\bibinfo {volume} {107}},\ \bibinfo
  {pages} {083510} (\bibinfo {year} {2023}{\natexlab{b}})},\ \Eprint
  {http://arxiv.org/abs/2212.00560} {arXiv:2212.00560 [hep-ph]} \BibitemShut
  {NoStop}%
\bibitem [{\citenamefont {Giblin}\ and\ \citenamefont
  {Tishue}(2019)}]{Giblin:2019nuv}%
  \BibitemOpen
  \bibfield  {author} {\bibinfo {author} {\bibfnamefont {J.~T.}\ \bibnamefont
  {Giblin}}\ and\ \bibinfo {author} {\bibfnamefont {A.~J.}\ \bibnamefont
  {Tishue}},\ }\href {\doibase 10.1103/PhysRevD.100.063543} {\bibfield
  {journal} {\bibinfo  {journal} {Phys. Rev. D}\ }\textbf {\bibinfo {volume}
  {100}},\ \bibinfo {pages} {063543} (\bibinfo {year} {2019})},\ \Eprint
  {http://arxiv.org/abs/1907.10601} {arXiv:1907.10601 [gr-qc]} \BibitemShut
  {NoStop}%
\bibitem [{\citenamefont {Kou}\ \emph {et~al.}(2021)\citenamefont {Kou},
  \citenamefont {Tian},\ and\ \citenamefont {Zhou}}]{Kou:2019bbc}%
  \BibitemOpen
  \bibfield  {author} {\bibinfo {author} {\bibfnamefont {X.-X.}\ \bibnamefont
  {Kou}}, \bibinfo {author} {\bibfnamefont {C.}~\bibnamefont {Tian}}, \ and\
  \bibinfo {author} {\bibfnamefont {S.-Y.}\ \bibnamefont {Zhou}},\ }\href
  {\doibase 10.1088/1361-6382/abd09f} {\bibfield  {journal} {\bibinfo
  {journal} {Class. Quant. Grav.}\ }\textbf {\bibinfo {volume} {38}},\ \bibinfo
  {pages} {045005} (\bibinfo {year} {2021})},\ \Eprint
  {http://arxiv.org/abs/1912.09658} {arXiv:1912.09658 [gr-qc]} \BibitemShut
  {NoStop}%
\bibitem [{\citenamefont {Kou}\ \emph {et~al.}(2022)\citenamefont {Kou},
  \citenamefont {Mertens}, \citenamefont {Tian},\ and\ \citenamefont
  {Zhou}}]{Kou:2021bij}%
  \BibitemOpen
  \bibfield  {author} {\bibinfo {author} {\bibfnamefont {X.-X.}\ \bibnamefont
  {Kou}}, \bibinfo {author} {\bibfnamefont {J.~B.}\ \bibnamefont {Mertens}},
  \bibinfo {author} {\bibfnamefont {C.}~\bibnamefont {Tian}}, \ and\ \bibinfo
  {author} {\bibfnamefont {S.-Y.}\ \bibnamefont {Zhou}},\ }\href {\doibase
  10.1103/PhysRevD.105.123505} {\bibfield  {journal} {\bibinfo  {journal}
  {Phys. Rev. D}\ }\textbf {\bibinfo {volume} {105}},\ \bibinfo {pages}
  {123505} (\bibinfo {year} {2022})},\ \Eprint
  {http://arxiv.org/abs/2112.07626} {arXiv:2112.07626 [gr-qc]} \BibitemShut
  {NoStop}%
\bibitem [{\citenamefont {Aurrekoetxea}\ \emph {et~al.}(2023)\citenamefont
  {Aurrekoetxea}, \citenamefont {Clough},\ and\ \citenamefont
  {Muia}}]{Aurrekoetxea:2023jwd}%
  \BibitemOpen
  \bibfield  {author} {\bibinfo {author} {\bibfnamefont {J.~C.}\ \bibnamefont
  {Aurrekoetxea}}, \bibinfo {author} {\bibfnamefont {K.}~\bibnamefont
  {Clough}}, \ and\ \bibinfo {author} {\bibfnamefont {F.}~\bibnamefont
  {Muia}},\ }\href {\doibase 10.1103/PhysRevD.108.023501} {\bibfield  {journal}
  {\bibinfo  {journal} {Phys. Rev. D}\ }\textbf {\bibinfo {volume} {108}},\
  \bibinfo {pages} {023501} (\bibinfo {year} {2023})},\ \Eprint
  {http://arxiv.org/abs/2304.01673} {arXiv:2304.01673 [gr-qc]} \BibitemShut
  {NoStop}%
\bibitem [{\citenamefont {{Turk}}\ \emph {et~al.}(2011)\citenamefont {{Turk}},
  \citenamefont {{Smith}}, \citenamefont {{Oishi}}, \citenamefont {{Skory}},
  \citenamefont {{Skillman}}, \citenamefont {{Abel}},\ and\ \citenamefont
  {{Norman}}}]{yt}%
  \BibitemOpen
  \bibfield  {author} {\bibinfo {author} {\bibfnamefont {M.~J.}\ \bibnamefont
  {{Turk}}}, \bibinfo {author} {\bibfnamefont {B.~D.}\ \bibnamefont {{Smith}}},
  \bibinfo {author} {\bibfnamefont {J.~S.}\ \bibnamefont {{Oishi}}}, \bibinfo
  {author} {\bibfnamefont {S.}~\bibnamefont {{Skory}}}, \bibinfo {author}
  {\bibfnamefont {S.~W.}\ \bibnamefont {{Skillman}}}, \bibinfo {author}
  {\bibfnamefont {T.}~\bibnamefont {{Abel}}}, \ and\ \bibinfo {author}
  {\bibfnamefont {M.~L.}\ \bibnamefont {{Norman}}},\ }\href {\doibase
  10.1088/0067-0049/192/1/9} {\bibfield  {journal} {\bibinfo  {journal} {The
  Astrophysical Journal Supplement Series}\ }\textbf {\bibinfo {volume}
  {192}},\ \bibinfo {eid} {9} (\bibinfo {year} {2011})},\ \Eprint
  {http://arxiv.org/abs/1011.3514} {arXiv:1011.3514 [astro-ph.IM]} \BibitemShut
  {NoStop}%
\end{thebibliography}%

\end{document}